\documentclass[preprint,superscriptaddress,nofootinbib]{revtex4}

\usepackage{graphicx}
\usepackage{amssymb}
\usepackage{color}

\begin{document}

\title{An investigation of the SCOZA for narrow square-well 
potentials and in the sticky limit}

\author{D. Pini}
\affiliation{Dipartimento di Fisica, Universit\`a degli Studi di
Milano, Via Celoria 16, 20133 Milano, Italy}

\author{A. Parola}
\affiliation{Dipartimento di Fisica e Matematica, Universit\`a dell'Insubria,
Via Valleggio 11, 22100 Como, Italy}

\author{J. Colombo}
\affiliation{Dipartimento di Fisica e Matematica, Universit\`a dell'Insubria,
Via Valleggio 11, 22100 Como, Italy}

\author{L. Reatto}
\affiliation{Dipartimento di Fisica, Universit\`a degli Studi di
Milano, Via Celoria 16, 20133 Milano, Italy}

\begin{abstract}

We present a study of the self consistent Ornstein-Zernike approximation 
(SCOZA) for square-well (SW) potentials of narrow width $\delta$. The main 
purpose 
of this investigation is to elucidate whether in the limit $\delta\to 0$, the 
SCOZA predicts a finite value for the second virial coefficient at the critical
temperature $B_{2}(T_{c})$, and 
whether this theory can lead to an improvement of the approximate Percus-Yevick 
solution of the sticky hard-sphere (SHS) model due to Baxter  
[R.~J.~Baxter, J. Chem. Phys. {\bf 49}, 2770 (1968)]. For SW of 
non vanishing $\delta$, the difficulties due to the influence of the 
boundary condition at high density already encountered in an earlier 
investigation [E.~Sch\"oll-Paschinger, A.~L.~Benavides, and 
R.~Casta\~neda-Priego, J. Chem.  Phys. {\bf 123}, 234513 (2005)] prevented us 
from obtaining reliable results for $\delta<0.1$. In the sticky limit this 
difficulty can be circumvented, but then the SCOZA fails to predict a 
liquid-vapor transition. The picture that emerges from this study is that 
for $\delta\to 0$, the SCOZA does not fulfill the expected prediction of a 
constant $B_{2}(T_{c})$ [M.~G.~Noro and D.~Frenkel, J. Chem. Phys. {\bf 113}, 
2941 (2000)], and that for thermodynamic
consistency to be usefully exploited in this regime, one should probably go 
beyond the Ornstein-Zernike {\em ansatz}.  

\end{abstract}

\maketitle

\section{Introduction}
\label{sec:intro}

Narrow attractive interactions are ubiquitous in the realm of colloidal 
dispersions. A most relevant instance is represented by depletion-induced 
forces in mixtures of highly asymmetric hard spheres, a field that has received
a fundamental contribution by Bob Evans~\cite{evans}. Although these 
investigations have 
shown that the structure of effective depletion interactions is in general more
complex than just a single attractive tail, a narrow attractive contribution at
short distance is still to be expected in most cases, and
its relevance to the modelization of both the equilibrium 
and non equilibrium phase behavior of colloidal systems has long been 
acknowledged. 

In view of this, it is somewhat disappointing that the difficulties 
experienced by most liquid-state theories when phase separation
occurs, the foremost of which is lack of convergence in a (not necessarily 
small) neighborhood of the critical region, only become worse when 
the range of the attractive potential decreases to a small fraction of the 
particle size. However, it may be argued that this is not a fundamental 
difficulty, inasmuch as this class of interactions can be dealt with by 
the sticky hard-sphere (SHS) model~\cite{baxter}. This is obtained from the  
prototypical attractive interaction, i.e., the hard-core plus square-well 
(SW) potential $v_{\rm SW}(r)$ given by 
\begin{equation}
v_{\rm SW}(r)=\left \{ \begin{array}{ll}
 +\infty \mbox{\hspace{1cm}}   & r < \sigma \, , \\
 -\epsilon & \sigma\leq r\leq (1+\delta)\sigma \, , \\
 0 & r>(1+\delta)\sigma \, ,
\end{array}
\right.
\label{sw}
\end{equation}
by taking the limits $\delta\to 0$, $\epsilon\to\infty$ so that
the second virial coefficient $B_{2}$ stays finite and non vanishing:  
\begin{equation}
\beta v_{\rm SHS}(r)=\left \{ \begin{array}{ll}
 +\infty \mbox{\hspace{1cm}}   & r < \sigma \, , \\
 -\ln\left[\displaystyle{\frac{1}{12\tau}}(1+\displaystyle{\frac{1}{\delta}}
)\right] & 
\sigma \leq r \leq (1+\delta)\sigma \mbox{\hspace{1cm}} \delta\to 0 \, , \\
 0 & r>(1+\delta)\sigma \, .
\end{array}
\right.
\label{shs}
\end{equation}
The stickiness parameter $\tau$ is related to the second virial coefficient of
the SHS model $B_{2}^{\rm SHS}$ by the relation
\begin{equation}
\tau=\frac{B_{2}^{\rm HS}}{4\, (B_{2}^{\rm HS}-B_{2}^{\rm SHS})} \, , 
\label{tau}
\end{equation}
where 
$B_{2}^{\rm HS}=2\pi\sigma^{3}/3$ is the second virial coefficient of the 
hard-sphere fluid. As shown by Baxter~\cite{baxter}, this model be can solved 
exactly within the Percus-Yevick (PY) integral equation~\cite{hansen}. 
Because of the way the limit~(\ref{shs}) is taken, the solution does not depend
on the temperature, but rather on the stickiness $\tau$ or, equivalently, on 
$B_{2}^{\rm SHS}$. The connection between the SHS model and an actual fluid 
at a certain temperature $T$ can be made by 
a generalized principle of corresponding states that, although not rigorous,
has been found to work remarkably well.
According to this principle, usually referred to as Noro-Frenkel (NF)
scaling~\cite{noro}, a pair potential $v(r)$ consisting of a hard-core
repulsion with diameter $\sigma$ and an attractive tail $w(r)$ can be mapped
into the SW potential~(\ref{sw}), provided that
the width $\delta$ of the attractive well is determined so                 
that its second virial coefficient $B_{2}(T)$ will be the same as that of the
original interaction at the same temperature $T$, assuming that in both cases
$T$ is measured in units of the well depth. On the other hand, when $\delta$ is
small, by following the same procedure the SW potential can in turn
be mapped into the SHS model.
In this way,
NF scaling can be combined with Baxter analytical solution of the SHS
model to provide a
unified description of fluids with narrow attractive interactions. For
instance, the critical temperature $T_{c}$ of such a fluid can be evaluated by 
that at
which its $B_{2}(T)$ is equal to the critical value of $B_{2}^{\rm SHS}$.

Although the SHS model has a perfectly regular solution within the PY 
approximation and has also been the subject of intensive simulation 
studies~\cite{seaton,kranendonk,miller,miller2}, the singularity embedded into 
its very definition should instigate a mild feeling of suspicion. In fact, 
twenty-three years after the appearance of the model, Stell~\cite{stell} 
discovered that the latter from a rigorous statistical mechanical point of view
is thermodynamically unstable whenever the number of particles is greater than
eleven. However, the state of affairs is less serious than it might seem: the
instability arises from a number of rather fancy particle configurations, that
are readily destroyed by the slightest amount of polidispersity in the system. 
Therefore, the SHS model and its approximate analytical solution via the PY 
equation~\cite{baxter} still remain a
valuable tool for the interpretation of the phase diagram of colloidal 
dispersions subject to short-range interactions.  

There is, however, a significant limitation of this solution that negatively 
affects its predictive power, i.e., its lack of thermodynamic consistency. This
is particularly evident when the liquid-vapor transition of the model is
considered: the coexistence curves determined by the compressibility and 
internal energy routes described in Sec.~\ref{sec:theory} differ markedly from
each other, the discrepancy between the critical densities amounting to nearly
a factor three, and neither of them agrees quantitatively with the simulation 
results~\cite{miller}. In this respect, it is also worth recalling that the PY
approximation upon which the analytical solution of the SHS is based, is 
expected to become less and less accurate as the density is increased, while
on the other hand the critical density increases as the range of the
interaction decreases, and for $\delta\to 0$ the limiting value predicted by 
the simulations~\cite{miller,largo} is considerably larger than that expected
for Lennard-Jones like fluids. 

Recently, a different approach has emerged, which is based upon enforcing 
consistency between the aforementioned energy and compressibility routes to the
thermodynamics, namely, the self-consistent Ornstein-Zernike 
approximation (SCOZA)~\cite{scozayuk1,scozayuk2,scozayuk4,foffi,scozaorea}. 
This has been applied especially to the case in which 
$v(r)$ consists of a hard core and a Yukawa (HCY) attractive tail: 
\begin{equation}
v_{\rm HCY}(r)=\left \{ \begin{array}{ll}
 +\infty \mbox{\hspace{1cm}}   & r < \sigma \, , \\
 -\epsilon \, \displaystyle{\frac{\sigma}{r}} \, e^{-z(r/\sigma-1)} & r>\sigma 
\, , 
\end{array}
\right.
\label{yuk}
\end{equation}
where $z$ is the inverse-range parameter of the interaction.
For the HCY potential, the SCOZA has shown
a remarkable robustness even for very short-range tails, while still remaining 
accurate~\cite{foffi,scozaorea}. 
On the other hand, one might argue that accuracy   
cannot be expected to persist for attractions of arbitrarily short 
range, since after all, as recalled in Sec.~\ref{sec:theory}, the SCOZA 
assumes that the direct correlation function $c(r)$ of the fluid is 
linear in the attractive part $w(r)$ of the interaction,  
and its low-density expansion does not even give back the correct 
expression of $B_{2}(T)$. In~\cite{orpanl} it was suggested that for narrow 
attractive interactions it could be more appropriate to have a $c(r)$ which 
depends linearly not on $w(r)$, but rather on its Mayer function 
$e^{-\beta w(r)}-1$, where $\beta=1/(k_{\rm B}T)$ is the inverse temperature. 
If this prescription were adopted in the SCOZA, one would obtain a ``non-linear
SCOZA'' bound to become exact at low density. 

Now, the SW potential is quite special in this respect, because both $w(r)$ 
and its
Mayer function are constant in the well region and vanish elsewhere, and 
therefore they have the same functional form as far as they dependence on $r$
is concerned, the only difference being their (state-dependent) amplitude. On 
the other hand, in the SCOZA this amplitude is determined by an {\em exact} 
thermodynamic consistency condition. As a consequence, for the SW potential 
the usual SCOZA {\em must} coincide with its non-linear counterpart, and hence
predict the exact $B_{2}(T)$ at low density, unlike for a generic tail 
potential. It is then tempting to investigate how the SCOZA will behave for 
narrow
SW potentials: will the quite special status of this potential enable it to 
recover a finite $B_{2}(T_{c})$ in the limit $\delta\to 0$? This possibility 
is quite enticing, because it would allow one to fix the lack of thermodynamic 
consistency that is the main defect of the PY solution of the SHS 
fluid~\cite{baxter}, 
possibly obtaining a better agreement with the simulation results for the phase
diagram~\cite{miller}. 

With this in mind, we have performed a study of the SCOZA for narrow SW 
interactions. Such a task requires a fully numerical solution of the SCOZA, 
because one cannot take advantage of the wealth of analytical results available
for the HCY potentials~\cite{hoye}. In fact, this problem was already tackled 
in~\cite{paschinger}, where the numerical solution of the SCOZA for a generic 
tail interaction $w(r)$ was worked out and applied to the SW potential. The 
investigation performed there showed that on narrowing the well, the results 
become more and more dependent on the choice of the upper boundary $\rho_{0}$ 
of the density interval where the SCOZA is solved, at which a high-density 
boundary condition needs to be specified. In order to get rid of such a 
dependence, one needs to move $\rho_{0}$ to higher and higher values, but 
eventually this is prevented by a lack of convergence of the numerical 
algorithm. On the basis of the conclusions of that work, we have then 
reconsidered the numerical solution of the SCOZA paying special attention to 
the problem of obtaining convergence at high density, and developed a numerical
algorithm that remains convergent up to $\rho_{0}\simeq 1.4\sigma^{-3}$, which 
is near the density at close packing. By this, we have managed to obtain 
results for wells down to $\delta=0.1$, which however 
is not small enough to represent reliably 
the behavior of the SCOZA in the limit $\delta\to 0$. 
     
We have then resorted to a different strategy, namely, we have taken the sticky
limit~(\ref{shs}) of the SW potential from the outset, and imposed 
thermodynamic consistency on the top of this. Again, the results turn
out to be strongly dependent on the high-density boundary condition, but in 
this case the simpler form of the SCOZA equation allows one to circumvent this 
problem by resorting to a solution technique which requires only the 
behavior at low density to be specified. Unfortunately, according to this 
solution the SCOZA fails to have a liquid-vapor transition in the sticky 
limit~(\ref{shs}). 

However disappointing, the full picture appears to be 
consistent enough to indicate that the SCOZA  
critical temperature will decrease too rapidly to give a finite value of 
$B_{2}(T_{c})$ for $\delta\to 0$.
On the other hand, the approximation which is introduced in the SCOZA, namely,
that the contribution to the direct correlation function $c(r)$ due to the tail
interaction $w(r)$ has the same range as $w(r)$ itself, and therefore vanishes
wherever $w(r)$ does; is made also in the PY equation, which moreover is 
thermodynamically inconsistent. This suggests that using thermodynamic 
consistency as a way to improve the PY solution for 
the SHS potential~(\ref{shs}) would require to give up the {\em ansatz} on 
the range of $c(r)$.  

The plan of the paper is as follows: in Sec.~\ref{sec:theory},  
the algorithm employed for the fully numerical solution in the 
case of a generic attractive tail $w(r)$ is 
described; in Sec.~\ref{sec:results} our results for SW potentials of several
amplitudes $\delta$ are presented, and compared with numerical simulations; 
the sticky limit of the SCOZA is considered in Sec.~\ref{sec:sticky}; our 
conclusions are presented in Sec.~\ref{sec:conclusions}. At the end of the 
paper, three Appendixes deal with some technical details pertinent to 
Sec.~\ref{sec:theory}: Appendix~\ref{sec:appint} presents the finite-difference
algorithm used to integrate the SCOZA partial differential equation (PDE); 
Appendix~\ref{sec:appopt} concerns the minimization algorithm which we used to
optimize $c(r)$ inside the repulsive core of the interaction; 
Appendix~\ref{sec:appcorr} describes the technique adopted for the treatment of
the hard-sphere part of the correlations at high density.

\section{The SCOZA for a generic attractive tail potential}
\label{sec:theory}

We are interested in a simple model fluid of particles interacting via 
a two-body, spherically symmetric potential $v(r)$ that consists of a hard core
with diameter $\sigma$ and an attractive tail $w(r)$:
\begin{equation}
v(r)=\left \{ \begin{array}{ll}
 +\infty \mbox{\hspace{1cm}}   & r < \sigma \, , \\
 w(r) & r>\sigma \, .
\end{array}
\right.
\label{pot}
\end{equation}
As is customary in integral-equation theories, the SCOZA is based on an 
approximate closure of the exact Ornstein-Zernike (OZ) equation that relates
the two-body correlation function $h(r)$ and the direct correlation function
$c(r)$:
\begin{equation}
h(r)=c(r)+\rho\!\int\!\!d^{3}{\bf r}^{\prime} \, c(|{\bf r}-{\bf r}^{\prime}|) 
\, h(r^{\prime})  \, ,
\label{oz}
\end{equation}
where $\rho\equiv N/V$ is the number density of the fluid, $V$ being the 
volume, and $N$ the particle number. The SCOZA closure reads
\begin{equation}
\left\{
\begin{array}{ll}
h(r)=-1                     & \mbox{$r<\sigma$} \, , \\
c(r)=c_{\rm HS}(r)-K(\rho, \beta) \, w(r) 
\mbox{\hspace{0.2cm}} & \mbox{$r>\sigma$}  \, ,
\end{array}
\right.
\label{closure}
\end{equation}
where $K(\rho, \beta)$ is a state-dependent function of the density $\rho$ and 
the inverse temperature $\beta=1/(k_{\rm B}T)$, and $c_{\rm HS}(r)$ is the 
direct correlation function of the hard-sphere fluid. The latter is assumed 
to be known, e.g. by the Verlet-Weis~\cite{verlet} or the 
Waisman~\cite{waisman} parametrization. Here the Waisman parametrization
has been used. Equation~(\ref{closure}) is similar to the well-known optimized 
random phase approximation (ORPA)~\cite{hansen}: in particular, the requirement
on $h(r)$ for $r<\sigma$, usually referred to as the core condition, is exact
because of the hard core of the potential~(\ref{pot}), while the expression of
$c(r)$ is clearly an approximation, since the off-core contribution to $c(r)$ 
is assumed to depend linearly on the interaction $w(r)$, hence having the 
same range as $w(r)$ itself --- the so-called OZ {\em ansatz}. 
However, in the ORPA
the amplitude of $w(r)$ is set to $K\!=\!\beta$, while in the SCOZA $K$ is 
{\em a priori} unknown, and must be determined so as to achieve consistency
between the compressibility and internal energy routes to the thermodynamics.
This requirement is expressed by the following condition on the isothermal
reduced compressibility $\chi_{\rm red}$ and the excess internal energy per 
particle $U$:
\begin{equation}
\frac{\partial}{\partial \beta}\left(\frac{1}{\chi_{\it red}}\right)=
\rho \frac{\partial^2 (\rho U)}{\partial \rho^2} \, ,
\label{consist}
\end{equation}
where it is understood that $\chi_{\rm red}$ and $U$ are determined 
respectively by the compressibility and internal energy routes, namely:
\begin{eqnarray}
\frac{1}{\, \chi_{\rm red}} & = & 1-\rho \!\int\!\! d^{3}{\bf r} \, c(r) \, ,
\label{comp} \\
U & = & \frac{1}{2}\, \rho \!\int\!\! d^{3}{\bf r} \, [h(r)+1] w(r) \, .
\label{energy}
\end{eqnarray}
Equation~(\ref{consist}) would in fact be a trivial identity in a hypothetical
exact description of the system, but this is not true anymore once an 
approximate closure such as Eq.~(\ref{closure}) is introduced. Together with 
Eqs.~(\ref{closure}), (\ref{comp}), (\ref{energy}), Eq.~(\ref{consist}) yields 
a PDE for $K(\rho, \beta)$, that must be solved
numerically. 
As in~\cite{scozayuk2,paschinger},
Eq.~(\ref{consist}) is rewritten using the internal energy per unit volume 
$u=\rho U$ as the unknown function:
\begin{equation}
D(\rho, u)\, \frac{\partial u}{\partial\beta}=\rho\frac{\partial^{2}u}
{\partial\rho^{2}} \, ,
\label{pde}
\end{equation}
where we have set  
\begin{equation}
D(\rho, u)=\frac{\partial}{\partial u}\left(\frac{1}{\, \chi_{\rm red}}
\right)_{\!\rho} \, .
\label{diff}
\end{equation}
The finite-difference algorithm used for the numerical integration of
Eq.~(\ref{pde}) has been described in detail in Appendix~\ref{sec:appint}.
Here we observe that, irrespective of the specific discretization of the PDE, 
a necessary step of the integration scheme consists in solving 
the OZ equation~(\ref{oz}) supplemented with the 
closure~(\ref{closure}). This amounts to solving the ORPA for an effective 
inverse temperature $\beta_{\rm eff}=K$. In several previous applications of 
the SCOZA, $w(r)$ was chosen as a Yukawa 
tail~\cite{scozayuk2,scozayuk1,scozayuk4,foffi}, a linear combination of Yukawa 
tails~\cite{scozayuk3,paschinger03}, or other similar forms such as the 
Sogami-ESE~\cite{paschinger04} or related potentials~\cite{yasutomi}. 
If $c_{\rm HS}(r)$ for $r>\sigma$ is assumed to be vanishing as in the 
PY equation for the hard-sphere fluid~\cite{hansen}, or is also 
given by a Yukawa 
tail as in the Waisman parametrization, these expressions of $w(r)$ allow for 
an almost
fully analytical solution of Eqs.~(\ref{oz}), (\ref{closure}). Although the
integration of the PDE must always be carried out numerically, this leads 
nevertheless to a considerable simplification of the whole procedure. 

For the
SW potential considered here, on the other hand, it is necessary to
solve also Eqs.~(\ref{oz}), (\ref{closure}) in a fully numerical fashion. 
Moreover, as already found in~\cite{paschinger} and recalled in the 
Introduction, for short-range wells the solution of the PDE is very sensitive
to the position of the upper boundary $\rho_{0}$ of the density interval, 
unless this is 
moved to very high values. Therefore, it is of paramount importance that the
numerical algorithm employed remains convergent in this regime. 
In~\cite{paschinger} Eqs.~(\ref{oz}), (\ref{closure}) were solved by the
Labik, Malijevsky, and Vonka algorithm~\cite{labik}. However, in that work 
it was found that the algorithm failed to converge at high density and low
temperature, so that for the shortest-range wells investigated there, namely
for $0.25 \leq \delta \leq 0.5$, the high-density boundary $\rho_{0}$ was not
moved beyond $\rho_{0}\sigma^{3}=1.15$. Therefore, here we have adopted a 
different method, which was originally developed in~\cite{pastore,pastore2} 
for the 
ORPA, and again applied to the ORPA for narrow SW and square-shoulder
potentials in~\cite{kahl}. 
This is based on the well-known property~\cite{hansen} that imposing the core 
condition in the ORPA is equivalent to requiring the Helmholtz free energy 
$A_{\rm RPA}$ given by the random phase approximation (RPA)~\cite{hansen} 
to be stationary with respect to
variations inside the hard core of the quantity $\phi(r)$ 
defined as 
\begin{equation}
\phi(r)\equiv c(r)-c_{\rm HS}(r) \, .
\label{phi}
\end{equation} 
We recall that 
outside the core one has $\phi(r)=-\beta w(r)$ in both the RPA and the ORPA. 
Specifically, the expression of $A_{\rm RPA}$ is   
\begin{equation}
-\, \frac{\beta A_{\rm RPA}}{N} = -\, \frac{\beta A_{\rm HS}}{N} + f_{\rm HTA} 
+ \frac{1}{2} f_{\rm Ring} \, ,
\label{expansion}
\end{equation}
where $A_{\rm HS}$ is the Helmholtz free energy of the hard-sphere fluid, and
$f_{\rm HTA}$ and $f_{\rm Ring}$ are given by 
\begin{eqnarray}
f_{\rm HTA} & = & \frac{1}{2}\, \rho \!\int\!\! d^{3}{\bf r}\, \phi(r)
[1+h_{\rm HS}(r)]
\label{fhta} \\
f_{\rm Ring} & = & -\frac{1}{\rho} \int\!\!\frac{d^{3}{\bf k}}{(2\pi)^{3}} \, 
\left\{\rho S_{\rm HS}(k)\hat{\phi}(k) + \ln\left[1-\rho S_{\rm HS}(k)
\hat{\phi}(k)\right]\right\} \, ,
\label{fring}
\end{eqnarray}
where $h_{\rm HS}(r)$ and $S_{\rm HS}(k)$ are respectively the two-body 
correlation function and the structure factor of the hard-sphere fluid, and  
the hats denote Fourier transforms. In the equations above, $f_{\rm HTA}$ 
refers to to the high-temperature approximation, whereby the two-body 
correlation function $h(r)$ is replaced by that of the hard-sphere fluid, while
$f_{\rm Ring}$ refers to the sum of the ring diagrams in the expansion of 
$A_{\rm RPA}$ in powers of $\phi(r)$~\cite{hansen}. If $f_{\rm Ring}$ 
is regarded as a functional of $\phi(r)$ inside the hard core, one finds
\begin{eqnarray}
\frac{\delta f_{\rm Ring}}{\delta\phi(r)} & = & \rho\, \Delta h (r) \, ,  
\label{first} \\
\frac{\delta^{2} f_{\rm Ring}}{\delta\phi(r)\delta\phi(r^{\prime})} & = &
\rho \!\int\!\!\frac{d^{3}{\bf k}}{(2\pi)^{3}} \, e^{i{\bf k}\cdot 
({\bf r}-{\bf r}^{\prime})} S^{\, 2}(k) \, 
\label{second}
\end{eqnarray}
where we have set $\Delta h(r)=h(r)-h_{\rm HS}(r)$, and $S(k)$ is the RPA 
structure factor given by 
\begin{equation}
S(k)=\frac{S_{\rm HS}(k)}{1-\rho S_{\rm HS}(k)\, \hat{\!\phi}(k)} \, . 
\label{srpa}
\end{equation}
Note that, since 
$h_{\rm HS}(r)$ satisfies exactly the core condition, $f_{\rm HTA}$ does not
depend on $\phi(r)$ for $r<\sigma$, and therefore does not contribute to the 
functional derivatives of $A_{\rm RPA}$ for $r<\sigma$. Equation (\ref{first})
shows that the core condition is satisfied if and only if the RPA 
free energy is stationary with respect to variations of $\phi(r)$ inside the 
core, and Eq.~(\ref{second}) shows that $f_{\rm Ring}$ is convex, 
so that this stationary point is unique and corresponds to a minimum of 
$f_{\rm Ring}$. 

As a consequence, one can solve the ORPA by minimizing $f_{\rm Ring}$ 
numerically with respect to $\phi(r)$ in the core region~\cite{pastore}. 
Since Eqs.~(\ref{first}),
(\ref{second}) hold irrespective of the form of $\phi(r)$ outside the core, 
this procedure can be equally well applied to the present case where 
(see Eq.~(\ref{closure})) $\phi(r)$ for $r>\sigma$ is given by 
$-K(\rho, \beta) w(r)$. Therefore, in order to solve the OZ equation~(\ref{oz})
with the closure~(\ref{closure}), we can, for any given $K$, minimize the 
functional $f_{\rm Ring}$ of Eq.~(\ref{fring}) with $\phi(r)=-K w(r)$ for 
$r>\sigma$.  The difference with respect to the ORPA consists in the fact that  
in the ORPA, the minimum of the functional gives the Helmholtz free energy of 
the system as predicted by the energy route, while this is not true anymore for
a generic $K$. In fact, one finds
\begin{equation}
\frac{\partial f_{\rm Ring}}{\partial K} = -\rho\! \int \!\! d^{3}{\bf r}\, 
\Delta h(r) w(r) = -2\, (U-U_{\rm HTA}) \, ,
\label{kfirst}
\end{equation}
where $U$ is given by Eq.~(\ref{energy}), and $U_{\rm HTA}$ is the 
corresponding quantity in the high-temperature approximation:
\begin{equation}
U_{\rm HTA}=\frac{1}{2}\, \rho \!\int\!\! d^{3}{\bf r} \, [h_{\rm HS}(r)+1] 
w(r) \, .
\label{uhta}
\end{equation}
For $K=\beta$, Eq.~(\ref{kfirst}) coincides with the usual thermodynamic 
relation $\partial(\beta A/N)/\partial\beta=U$ with $A$ given by 
Eq.~(\ref{expansion}), so that $A$ is indeed the Helmholtz free energy obtained
by integrating the internal energy $U$ with respect to $\beta$. On the other
hand, this is not the case if $K$ has a nontrivial dependence on $\beta$. 
Therefore, unlike in the ORPA and the RPA, in the 
SCOZA Eqs.~(\ref{expansion}), (\ref{fhta}), (\ref{fring}) do not give 
the Helmholtz free energy. Nevertheless, Eq.~(\ref{kfirst}) turns out to be 
very useful to overcome a difficulty related to the numerical solution of the
SCOZA. Specifically, the calculation of the quantity $D(\rho ,u)$ defined in 
Eq.~(\ref{diff}) requires that Eqs.~(\ref{oz}), 
(\ref{closure}) must be solved at fixed $\rho$ and $U$, rather than at fixed
$\rho$ and $K$, i.e., for any given density one needs to find the value of $K$ 
such that the internal energy per particle takes a given value $U=\bar{U}$. One way to
do this would be to change $K$ by trial and error, solve Eqs.~(\ref{oz}), 
(\ref{closure}) with respect to $\phi(r)$ inside the core for each $K$, and 
find the corresponding $U$ by Eq.~(\ref{energy}), until
the condition $U=\bar{U}$ is met up to a prescribed accuracy. 
However, we found that one can solve Eqs.~(\ref{oz}), (\ref{closure}) 
simultaneously with respect to $\phi(r)$ and $K$ in a single optimization run.
To this end, let us observe that one has
\begin{equation}
\frac{\partial^{2} f_{\rm Ring}}{\partial K^{2}} = \frac{\rho}{2}
\int\!\!\frac{d^{3}{\bf k}}{(2\pi)^{3}} \, \left[S(k) w(k)\right]^{2} \, ,
\label{ksecond}
\end{equation}
where $S(k)$ is given by Eq.~(\ref{srpa}).
We now introduce the Legendre transform of $f_{\rm Ring}$ 
\begin{equation}
{\cal S} = f_{\rm Ring}-K \frac{\partial f_{\rm Ring}}{\partial K} = 
f_{\rm Ring} + 2K (U-U_{\rm HTA}) \, .
\label{s}
\end{equation}
In the case of the ORPA in which $K=\beta$, ${\cal S}$ coincides with twice the
excess entropy per particle with respect to the hard-sphere fluid. Equations
(\ref{kfirst}), (\ref{ksecond}) imply that, if $2(U-U_{\rm HTA})$ is fixed at 
some given value $\Delta U$, ${\cal S}$ is a convex function of $K$ such that
\begin{equation}
\frac{\partial {\cal S}}{\partial K} = -\rho\!\int\!\!d^{3}{\bf r}\, 
\Delta h(r) w(r) + \Delta U \, ,
\label{sfirst}
\end{equation}
so that its minimum is obtained for that $K$ which, when inserted into 
Eq.~(\ref{closure}), will yield $\Delta U$ via Eqs.~(\ref{energy}), 
(\ref{uhta}) for the internal energy. Therefore, in order to solve the OZ 
equation~(\ref{oz}) with the closure~(\ref{closure}) for a fixed density 
$\rho$ and internal energy $U$, it is sufficient to minimize ${\cal S}$, 
regarded as a function of $K$ and a functional of $\phi(r)$ for $r<\sigma$. 
The numerical procedure adopted for the minimization is described in detail
in Appendix~\ref{sec:appopt}.

A difficulty related to the aforementioned request that Eqs.~(\ref{oz}), 
(\ref{closure}) must be solved also at very high density, is that in this 
regime the structure factor of the fluid develops very high and narrow, 
``pseudo-Bragg'' peaks. These do not have a direct physical meaning, since they
occur in a density range where the stable phase would actually be the solid 
one. Nevertheless, they must be taken into account when solving Eqs.~(\ref{oz}),
(\ref{closure}) numerically. In fact, this problem appears already for the 
simple hard-sphere fluid: if one starts from the (analytical) 
$\hat{c}_{\rm HS}(k)$ given by the Waisman parametrization, finds 
$\hat{h}_{\rm HS}(k)$ by the OZ equation~(\ref{oz}) and obtains
$h_{\rm HS}(r)$ by performing directly the inverse Fourier transform of 
$\hat{h}_{\rm HS}(k)$, it is found that for $\rho^{*}\gtrsim 1.2$, the
core condition is very poorly satisfied. The reason for this is that many 
of the peaks of $\hat{h}_{\rm HS}(k)$ will simply be missed by the discrete 
sampling of the wave vector, unless the number of point used in the transform
is so large (about $2^{18}$), that the numerical computation becomes 
unpractical. In order to cope with this problem, the contribution of these 
peaks to the inverse Fourier transform has been determined analytically by 
the method described in Appendix~\ref{sec:appcorr}.   

The initial condition required by the PDE~(\ref{pde}) at $\beta=0$ and the 
boundary conditions at $\rho=0$ and on the spinodal curve, i.e., the locus of 
diverging compressibility below the critical temperature, have been detailed
elsewhere~\cite{scozayuk2}. 
The high-density boundary condition is more problematic in this context, since 
this is not known exactly while, as stated in Sec.~\ref{sec:intro}, the results for narrow SW are quite
sensitive to both the position of the high-density boundary $\rho_{0}$, and   
the kind of approximation used for $u$ at $\rho=\rho_{0}$.
This point will be considered in more detail in Sec.~\ref{sec:results}; 
here we remark that in this investigation the high-density boundary condition 
has not been imposed on the second derivative of $u$ with respect to the 
density $\partial^{2} u/\partial \rho^{2}$ as in a number of previous 
works~\cite{scozayuk2,scozayuk4,foffi,paschinger}, 
but rather directly on $u$, because we found that the results obtained in this
way were less sensitive to the choice of $u(\rho_{0}, \beta)$. 

The vapor-liquid coexistence curve is obtained
by imposing the conditions of thermodynamic equilibrium, i.e., by finding the
densities $\rho_{v}$, $\rho_{l}$ which give the same pressure $P$ and 
the same chemical potential $\mu$ on the liquid and vapor branches of the 
sub-critical isotherm. $P$ and $\mu$ are determined by integrating with respect
to $\beta$ the exact identities $\partial(\beta P)/\partial\beta=
-u+\rho\partial u/\partial\rho$, $\partial(\beta\mu)/\partial\beta=
\partial u/\partial\rho$, which are fulfilled also by the SCOZA because of the 
consistency condition~(\ref{consist}). 

Concerning the parameters used in the numerical calculation, we typically set 
the density step at $\Delta\rho^{*}=10^{-3}$, and the initial 
inverse-temperature step at $\Delta\beta_{0}^{*}=10^{-3}$. 
In order to locate the critical point accurately,
$\Delta\beta$ is decreased as the critical point is approached, and then 
gradually expanded back to $\Delta\beta_{0}$ below the critical temperature. 
We found that the results are not very sensitive to the density or temperature 
mesh, e.g., even a substantial coarsening of the mesh such as 
$\Delta\rho^{*}=10^{-2}$, $\Delta\beta_{0}^{*}=10^{-2}$ leaves 
the results nearly 
unaffected, except for the coexistence curve in the neighborhood of the 
critical point. In this case, a very high accuracy in the numerical calculation
is required to fulfill the conditions of thermodynamic equilibrium, so that a 
coarse mesh results in a large gap at the top of the coexistence curve. On the 
other hand, the numerical output turns out to be much more sensitive to the
discretization used in the numerical Fourier transform. We then used a rather 
small step in real space, $\Delta r^{*}=5\times 10^{-4}$, and a large number 
of mesh points, $N=2^{15}$. We employed the fast Fourier transform routines 
developed by Takuya Ooura at the Research Institute of Mathematical Sciences, 
Kyoto University, and made freely available online.  

Whenever the function to be transformed is discontinuous, such 
as $\Delta h(r)$ or $\Delta c(r)$, it is best to subtract off the 
discontinuity, and add the slowly decaying transform of the discontinuous 
part determined analytically at the end. Here we implemented this procedure on 
both the function and its derivative, so that the numerical transform is done
on a smooth function that is not only continuous, but also differentiable. If
$D\!f(r_{0})$, $D\!f^{\prime}(r_{0})$ denote the discontinuities respectively 
of $f$ and its derivative $f^{\prime}$ at $r\!=\! r_{0}$, i.e., 
$D\!f(r_{0})=\lim_{r\to r_{0}^{-}}f(r)-\lim_{r\to r_{0}^{+}}f(r)$, 
$D\!f^{\prime}(r_{0})=\lim_{r\to r_{0}^{-}}f^{\prime}(r)-\lim_{r\to r_{0}^{+}}
f^{\prime}(r)$, the function $\tilde{f}$ that is transformed numerically is 
given by
\begin{equation}
\tilde{f}(r)=f(r)-\left[D\!f(r_{0})+(r-r_{0})D\!f^{\prime}(r_{0})\right]
\left[1-\Theta(r-r_{0})\right] \, , 
\label{disc}
\end{equation}
where $\Theta(r)$ is the Heaviside step function defined by $\Theta(r)=1$ for 
$r>0$, $\Theta(r)=0$ for $r<0$. 
For the SW potential studied here, the discontinuities of $\Delta h(r)$ and 
$\Delta c(r)$ are located at $r_{0}=\sigma$ and $r_{0}=\sigma (1+\delta)$. 

The accuracy within which the core condition is fulfilled can be determined 
{\em a posteriori} by calculating the radial distribution function 
$g(r)\equiv h(r)+1$ inside the repulsive core $r<\sigma$. Typically, we found
that for $T\simeq T_{c}$ and high densities such as 
$\rho\simeq 0.9\sigma^{-3}$,
the order of magnitude of $g(r)$ is $g(r)\sim 10^{-8}$-$10^{-7}$ for 
$0.1\sigma \lesssim r<\sigma$ and $g(r)\sim 10^{-6}$-$10^{-3}$ for 
$r\lesssim 0.1\sigma$.

\section{Results}
\label{sec:results}

We have employed the numerical solution of the SCOZA described in 
Sec.~\ref{sec:theory} to study SW potentials of several widths $\delta$.  
For large $\delta$, our results are basically identical to those obtained
in~\cite{paschinger}. As can be inferred from the comparison between 
Tab.~\ref{tab:crit} shown further in this Section and Tab.~I 
of~\cite{paschinger}, the critical points for $\delta\geq 0.5$ differ at most 
by $\sim 0.2\%$. Here we discuss in detail our data for   
$\delta\leq 0.25$, which is the smallest value studied in~\cite{paschinger}. 
In this regime, the liquid-vapor transition considered here is actually 
expected to be metastable with respect to freezing~\cite{liu}. 

As already found in~\cite{paschinger} and anticipated in Secs.~\ref{sec:intro},
\ref{sec:theory}, we also have observed a strong influence on the results 
of the 
high-density boundary condition, that becomes more and more important as 
$\delta$ decreases. This is especially true for the critical density 
$\rho_{c}$ and the high-density branches of the spinodal and coexistence 
curves. In order to assess this sensitivity, the high-density boundary 
$\rho_{0}$ was initially set at $\rho_{0}^{*}=1$, and then moved forward to 
values as high as $\rho_{0}^{*}=1.4$, which is very close to the close-packing
density $\rho_{0}^{*}=\sqrt{2}$. Moreover, for each $\rho_{0}$ several 
different approaches for the boundary condition were tried. Specifically, 
the internal energy $u(\rho_{0}, \beta)$ has been determined 
by the HTA; the ORPA; the ``non-linear ORPA''~\cite{orpanl}, 
whereby $\beta$ is replaced by the Mayer function $e^{\beta^{*}}-1$ as the
amplitude of the potential in $c(r)$, i.e., one sets 
$K^{*}=e^{\beta^{*}}-1$ in Eq.~(\ref{closure}); the EXP~\cite{hansen}, 
which is
obtained by setting $g_{\rm EXP}(r)=g_{\rm HS}(r)\exp[\Delta h_{\rm ORPA}(r)]$,
where $g(r)$ is the radial distribution function, and 
$\Delta h_{\rm ORPA}(r)\!=\! h_{\rm ORPA}(r)-h_{\rm HS}(r)$; the linearized 
version of EXP known as LIN~\cite{hansen}, where one has 
$g_{\rm LIN}(r)=g_{\rm HS}(r)[1+\Delta h_{\rm ORPA}(r)]$; and, for very narrow 
wells, the solution of the PY equation in the SHS 
limit~\cite{baxter}. For the values of $\delta$ discussed
below, the results were found to be enough stable with respect to a further 
increase of $\rho_{0}$ or a change of the approximation used for $u(\rho_{0}, 
\beta)$ to warrant comparison with other predictions.

\begin{figure}
\includegraphics[width=10cm]{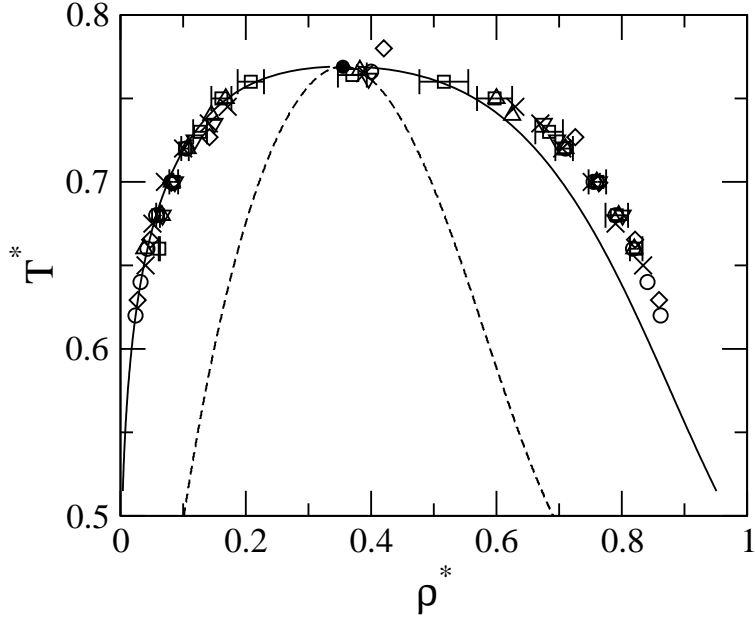}
\caption{Liquid-vapor coexistence curve and critical point of the SW 
fluid of width $\delta=0.25$ in the density-temperature plane. 
Solid line: SCOZA. Full circle: SCOZA critical point. Squares with error bars:
simulation data by Vega et al.~\cite{vega}. 
Diamonds: simulation data by Elliott et al.~\cite{hu}. 
Triangles: simulation data by Liu et al.~\cite{liu}
Inverted triangles: simulation data by del R\'\i o at al.~\cite{rio}. 
Circles: simulation data by Orea et al.~\cite{orea}. 
Crosses: simulation data by Pagan et al.~\cite{pagan}. 
The SCOZA spinodal has also been shown by the dashed 
line.}
\label{fig:025}
\end{figure}

In Fig.~\ref{fig:025} we have reported the SCOZA coexistence and spinodal 
curves for $\delta=0.25$ obtained with $\rho_{0}^{*}=1.2$ and 
$u(\rho_{0}, \beta)$ given by the non-linear ORPA. 
The critical temperature and density predicted by the
SCOZA are $T_{c}^{*}=0.769$, $\rho_{c}^{*}=0.355$. These values are close, but
not identical to those found in~\cite{paschinger} for the same value of 
$\delta$, namely $T_{c}^{*}=0.761$, $\rho_{c}^{*}=0.343$. The difference is 
to be traced back to both the somewhat lower value of $\rho_{0}$ used
there, $\rho_{0}^{*}=1.15$, and the fact that the boundary condition at 
$\rho_{0}$ was imposed on $\partial^{2} u/\partial\rho^{2}$ rather than $u$. 
As observed in Sec.~\ref{sec:theory}, we found that this enhanced the 
sensitivity of the results to the choice of the approximation used at 
$\rho_{0}$, namely, the HTA in the case of~\cite{paschinger}. The main 
difference between the two coexistence curves is that the liquid branch of the
present one is slightly shifted to higher densities with respect to that 
obtained in~\cite{paschinger}. Nevertheless, the two curves are close to each 
other, and the comparison with the simulation 
results~\cite{vega,hu,liu,rio,pagan,orea} shown in 
Fig.~\ref{fig:025} confirms the trend already found there, i.e., the liquid 
branch of the SCOZA coexistence curve underestimates the simulation results. 
The different simulations reported in the figure are enough consistent with one
another to indicate that this is indeed an inaccuracy of the SCOZA.  
The critical temperature is in good
agreement with most simulation results~\cite{vega,liu,rio,pagan,orea}, while 
the critical density is underestimated with respect to all of them, although it 
lies within the error-bar given in~\cite{vega}, and  
the discrepancy is generally smaller than that observed on the liquid branch 
of the coexistence curve. 

\begin{figure}
\includegraphics[width=10cm]{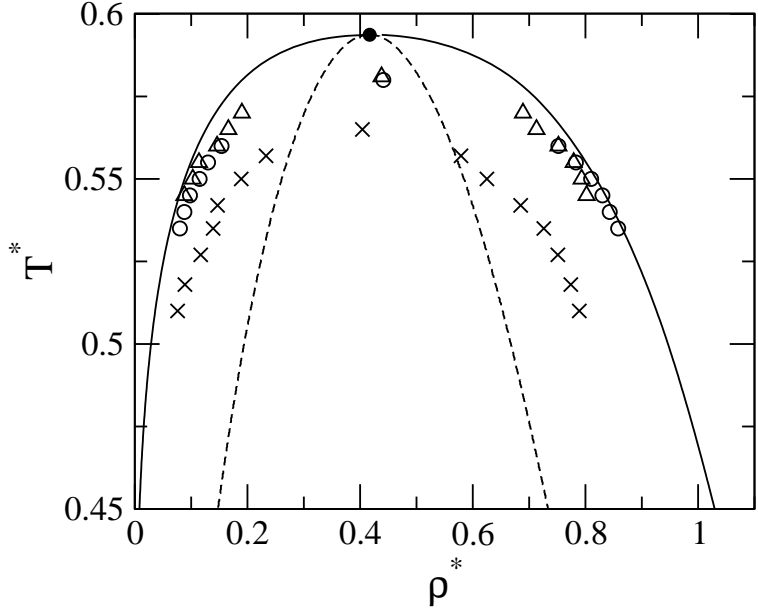}
\caption{Same as Fig.~\ref{fig:025} for $\delta=0.15$. Solid line: SCOZA. 
Full circle: SCOZA critical point. Circles: simulation data by Orea et 
al.~\cite{orea}. Crosses: simulation data by Pagan et al.~\cite{pagan}.  
Triangles: simulation data by Liu et al.~\cite{liu}. 
Dashed line: SCOZA spinodal.}  
\label{fig:015}
\end{figure}

\begin{figure}
\includegraphics[width=10cm]{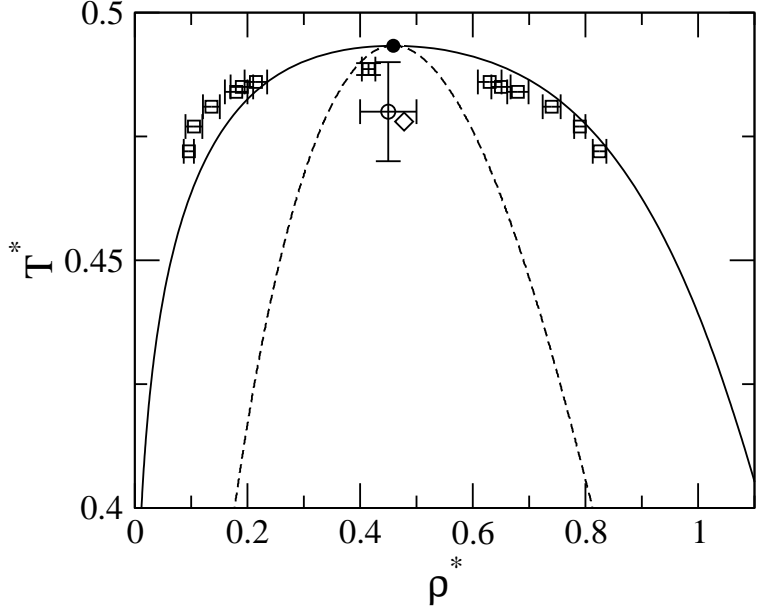}
\caption{Same as Fig.~\ref{fig:025} for $\delta=0.1$. Solid line: SCOZA. 
Full circle: SCOZA critical point. Squares with error bars: simulation data
by Duda~\cite{duda}. Circle with error bars: simulation critical point by 
Skibinsky et al.~\cite{skibinsky}. Diamond: simulation critical point by 
Largo et al.~\cite{largo}. Dashed line: SCOZA spinodal.}
\label{fig:01}
\end{figure}

Figures~\ref{fig:015}, 
\ref{fig:01} show the SCOZA spinodal and coexistence curves respectively for
$\delta=0.15$ and $\delta=0.1$, together with simulation data for 
the coexistence curve and the critical 
point~\cite{skibinsky,pagan,liu,orea,largo,duda}. 
The SCOZA results were again obtained using 
the non-linear ORPA as the boundary condition at $\rho_{0}$, and setting 
$\rho_{0}^{*}=1.4$. 
Unlike what found for $\delta=0.25$, the coexistence curve 
of the SCOZA is wider than
most simulation predictions. This behavior is largely due to the fact that now 
the SCOZA critical temperature is appreciably higher than the simulation ones. 

As a general remark, the overall agreement with the phase diagram predicted by 
the SCOZA appears to be worse than that found for HCY potentials of comparable 
range~\cite{scozayuk4,foffi,scozaorea}, although for $\delta=0.15$ and 
$\delta=0.1$ a quantitative assessment of the performance of the SCOZA
is difficult, because of the rather large discrepancies shown by different 
simulations. 
A perhaps more conclusive comparison can be made with the recent
simulation results for the critical point obtained in~\cite{largo}, some of
which have been reported in Tab.~\ref{tab:crit} together with the SCOZA 
predictions. These indicate that the SCOZA for the SW fluid 
underestimates $\rho_{c}$, the discrepancy on 
$\rho_{c}$ being larger than that on $T_{c}$. 
In a previous investigation of the SCOZA for narrow HCY 
fluids~\cite{scozaorea}, $\rho_{c}$ was again found to deviate from simulations
more than $T_{c}$, although in that case $\rho_{c}$ was 
overestimated by the SCOZA. 
\begin{table}
\begin{tabular}{lllllllll} \hline\hline
 & & \multicolumn{3}{c}{SCOZA} & & \multicolumn{3}{c}{MC} \\
 \cline{3-5} \cline{7-9}
\multicolumn{1}{c}{$\delta$} & & \multicolumn{1}{c}{$T_{c}^{*}$} & & \multicolumn{1}{c}{$\rho_{c}^{*}$} & & \multicolumn{1}{c}{$T_{c}^{*}$} & & \multicolumn{1}{c}{$\rho_{c}^{*}$} \\
 $0.8$ &   & $1.953$ &    & $0.247$ &    & $1.940$  &    & $0.263$        \\
 $0.5$ &   & $1.211$ &    & $0.272$ &    & $1.220$  &    & $0.310$        \\  
 $0.3$ &   & $0.850$ &    & $0.328$ &    & $0.847$  &    & $0.376$        \\  
 $0.2$ &   & $0.683$ &    & $0.382$ &    & $0.667$  &    & $0.421$        \\
 $0.1$ &   & $0.493$ &    & $0.459$ &    & $0.4780$ &    & $0.478$ \\ 
\hline\hline
\end{tabular}
\caption{Critical temperature and density of the SW fluid for several
values of the well width $\delta$, as obtained by the SCOZA and Monte Carlo 
(MC) numerical simulations performed in~\cite{largo}. According to the estimate
given in~\cite{largo} for the case $\delta=0.05$ (not reported here), the 
errors in the MC critical parameters are about $\pm 0.14\%$ for $T_{c}^{*}$ and 
$\pm 1.56\%$ for $\rho_{c}^{*}$.}
\label{tab:crit}
\end{table}

A qualitative feature common to the SCOZA phase diagrams shown in 
Figs.~\ref{fig:025}, \ref{fig:015}, \ref{fig:01} is that the coexistence curve
is much wider then the spinodal curve, so that even at densities near the 
critical one, and {\em a fortiori} at low or high densities, there is a sizable
temperature interval where the fluid may survive in a single phase as a 
metastable state, before spinodal decomposition sets in. According to this, in 
the experimental practice one should be circumspect before using spinodal 
decomposition to identify equilibrium phase separation on the assumption that 
the coexistence and spinodal curves run close to each other. In fact, according
to the SCOZA they do not, especially at the low volume fractions that are 
frequently considered in experiments on colloidal dispersions.  

For wells such that $\delta\leq 0.05$, the dependence of the SCOZA on the 
high-density boundary condition becomes so strong, that the results do not show
any clear trend towards saturation in the region $\rho_{0}^{*}\leq 1.4$. It is
possible that this could be achieved by moving $\rho_{0}$ to even higher 
values, bigger than that corresponding to the physical close packing. 
However, with the 
fully numerical algorithm considered here we failed to obtain solutions of 
Eqs.~(\ref{oz}), (\ref{closure}) beyond close packing, because in this regime
the correlation functions become very singular, and even the prescription 
described in Appendix~\ref{sec:appint} proved insufficient, at least for the 
number of points in the numerical Fourier transform that can be handled in 
practice. Therefore, the results that we have obtained for $\delta\leq 0.05$
cannot be considered meaningful. 

It is worthwhile pointing out that the sensitivity on the boundary condition at
high density is much stronger for the SW potential considered here than for 
the HCY potential. In order to compare the two cases, it is useful to resort
to the NF mapping~\cite{noro} mentioned in Sec.~\ref{sec:intro}, 
according to which a fluid of hard-core
particles of diameter $\sigma$ with a given attractive tail potential at 
reduced temperature $T^{*}$ can be mapped into a SW fluid with the
same $\sigma$, and a range determined so that its second virial coefficient 
$B_{2}(T)$ coincides with that of the original fluid for $T=T^{*}$. 
We remark that here this mapping is used without any assumption as to the 
behavior of the system for interactions of vanishing range, so its use is fully
legitimate also within the SCOZA. 
A HCY fluid with inverse-range parameter 
$z=5.5$ at $T^{*}=0.48$, which is the critical value according to the SCOZA, 
should then be equivalent to a SW fluid with $\delta=0.1$ at the same $T^{*}$. 
The SCOZA critical temperature of this SW,
$T_{c}^{*}=0.49$, is indeed close to that of the HCY, in agreement with the 
prediction from the mapping, while the difference between the critical 
densities, $\rho_{c}^{*}=0.459$ for the SW and $\rho_{c}^{*}=0.427$ 
for the HCY, although larger, is still below $10\%$.  
However, for the HCY fluid setting $\rho_{0}^{*}=1$ is sufficient to obtain 
a stable critical point, while for the SW $\rho_{0}^{*}$ must be set
at a substantially higher value, $\rho_{0}^{*}=1.4$.  

Further insight into the difference 
between the two potentials is obtained by checking the behavior of the 
effective range $\delta$ of the HCY below $T_{c}$: we find that, 
as the 
temperature decreases, the effective range also decreases. As a consequence,
for $T<T_{c}$ it should be possible to map the densities at coexistence 
of the HCY fluid into those of a SW fluid at the same $T^{*}$, but with 
a {\em lower} $T_{c}^{*}$, i.e., at a higher rescaled temperature 
$T/T_{c}$. This implies that, according to the NF mapping, the 
coexistence curve of the HCY fluid in the $T$-$\rho$ plane should be 
narrower than that of the ``equivalent'' SW at $T_{c}^{*}$. 
\begin{figure}
\includegraphics[width=10cm]{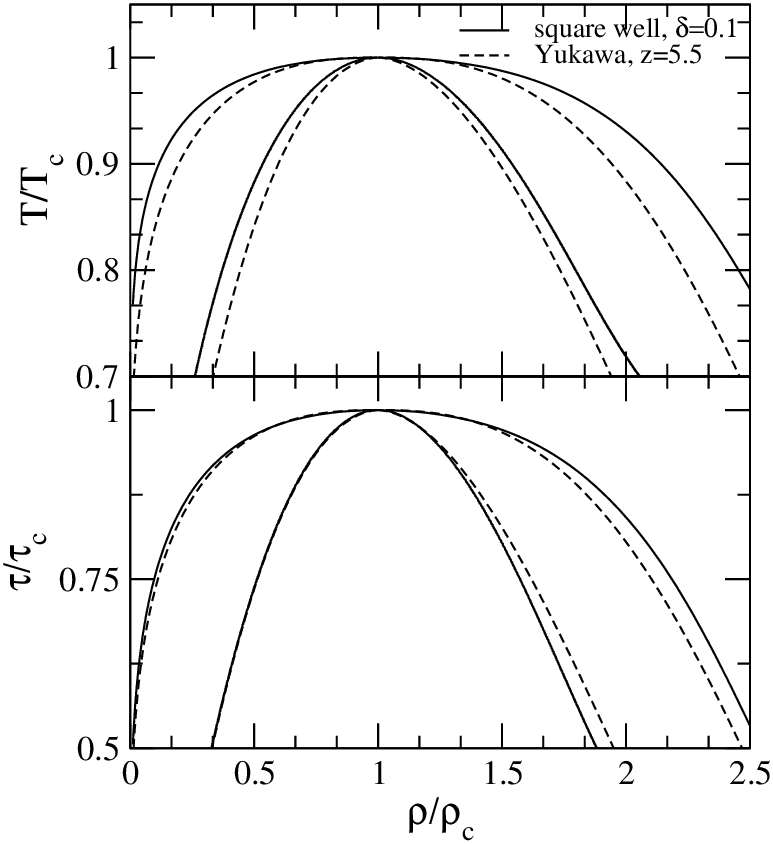}
\caption{Solid line: SCOZA coexistence (outer) and spinodal 
(inner) curves for 
a SW fluid of width $\delta=0.1$. Dashed line: same for a HCY fluid with 
inverse-range parameter $z=5.5$. The upper panel shows the curves in the 
temperature-density plane, with both variables rescaled by their critical 
values. In the lower panel the rescaled temperature has been replaced 
by the rescaled
stickiness parameter $\tau/\tau_{c}$~\cite{baxter}. Note how the curves for 
the two potentials get closer in the latter case.} 
\label{fig:coexred}
\end{figure}
That this is actually the case is shown in the upper panel of 
Fig.~\ref{fig:coexred}, where the
coexistence and spinodal curves of the HCY fluid with $z=5.5$ and the 
SW fluid with $\delta=0.1$ have been plotted using rescaled 
temperatures $T/T_{c}$ and densities $\rho/\rho_{c}$. 
It is evident that the
SW coexistence curve is appreciably wider, especially when the liquid
branch is considered. It is also worthwhile considering the behavior of the 
coexistence and spinodal curves when the temperature is replaced by $B_{2}(T)$,
or equivalently by the stickiness parameter $\tau$ obtained by 
replacing $B_{2}^{\rm SHS}$ with $B_{2}(T)$ in Eq.~(\ref{tau}). 
The NF mapping predicts that, with this choice of the
temperature-like variable, the phase diagram of narrow attractive potentials
should become universal. 
The two fluids considered here do give similar values for
$\tau$ at the critical point, namely, $\tau_{c}=0.115$ for the SW and 
$\tau_{c}=0.107$ for the HCY. If the rescaled temperature $T/T_{c}$ is
replaced by the rescaled stickiness $\tau/\tau_{c}$, as in the lower panel of
Fig.~\ref{fig:coexred}, the vapor branches of the coexistence and spinodal 
curves become nearly coincident, and the discrepancy between the liquid 
branches of the coexistence curve becomes substantially smaller, although the 
SW still gives a wider curve. Quite curiously, the situation with the
liquid branches of the spinodals is reversed with respect to that shown in the
upper panel, i.e., as a function of $\tau$ the SW spinodal is narrower
than the HCY spinodal, while as a function of $T$ it is wider. On the basis
of the above considerations, the sensitivity of the
SW potential to the high-density boundary condition may depend at 
least in part on the fact that its phase diagram is shifted towards higher
densities than those of a HCY fluid of equivalent range.    

\begin{figure}
\includegraphics[width=10cm]{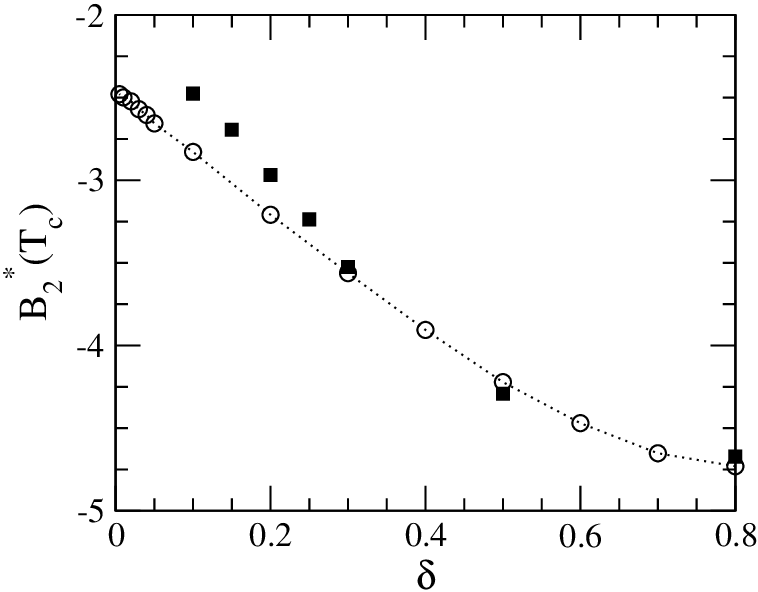}
\caption{Reduced second virial coefficient of the SW fluid at the critical 
temperature $B_{2}^{*}(T_{c})$ as a function of the well width $\delta$. 
Squares: SCOZA. Circles: simulation results by Largo et al.~\cite{largo}. 
The dotted line is a guide to the eye.}  
\label{fig:virial}
\end{figure}
The behavior of $B_{2}(T_{c})$, shown in Fig.~\ref{fig:virial}, 
is in semi 
quantitative agreement with the simulation results~\cite{largo}. In both cases,
the $B_{2}(T_{c})$ increases as $\delta$ decreases. However, because of the 
impossibility to obtain meaningful results for the critical
point of narrow wells such that $\delta\leq 0.05$,       
we are not able to say what the fate of the SCOZA $B_{2}(T_{c})$ 
will be in the limit $\delta\to 0$, namely, if it will tend to a finite
value as predicted by the simulations~\cite{miller,largo}. 
The arguments put forth in Sec.~\ref{sec:sticky} rather suggest that 
the SCOZA would not yield
a finite value of $B_{2}(T_{c})$ for vanishing attraction range.

\begin{figure}
\includegraphics[width=10cm]{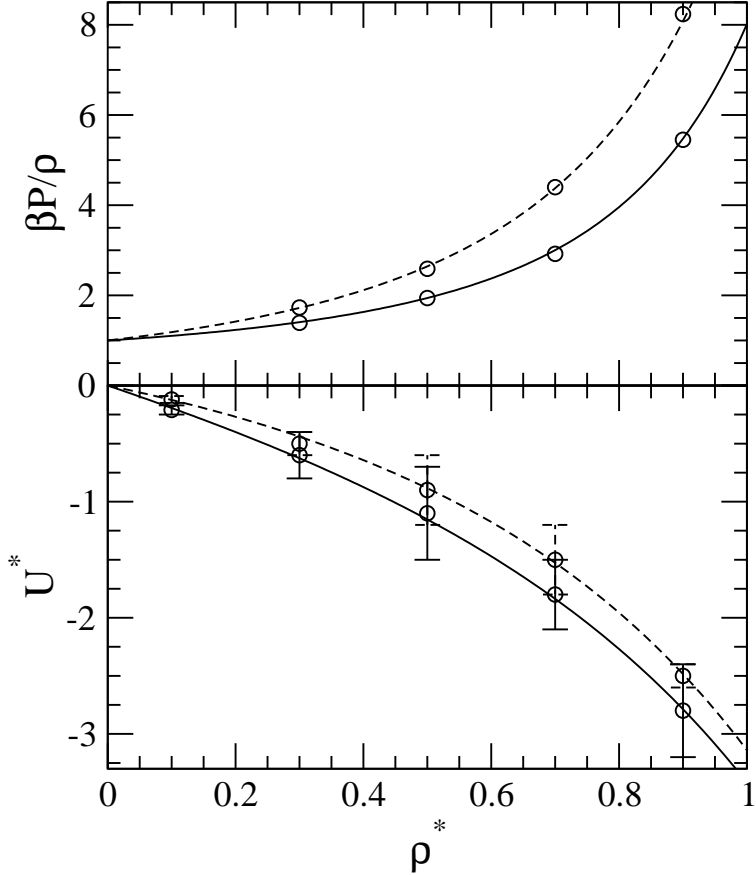}
\caption{Compressibility factor $\beta P/\rho$ (upper panel) and excess 
internal energy per particle $U^{*}$ (lower panel) of the SW fluid with 
$\delta=0.1$. 
Solid line: SCOZA for $T^{*}=1$. Dashed line: SCOZA for $T^{*}=2$. Circles: 
simulation results at the same temperatures~\cite{bergenholtz}.}   
\label{fig:zcomp}
\end{figure}

\begin{figure}
\includegraphics[width=10cm]{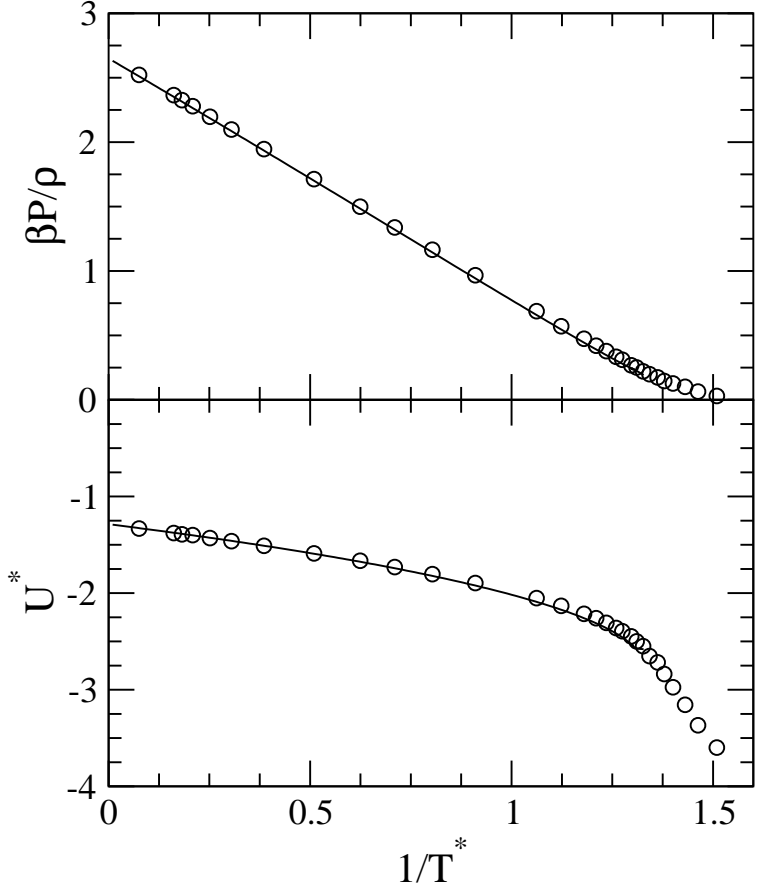} 
\caption{Compressibility factor $\beta P/\rho$ (upper panel) and excess
internal energy per particle $U^{*}$ (lower panel) of the SW fluid with 
$\delta=0.25$ and $\rho^{*}=0.42$. Solid line: SCOZA. Circles: simulation 
results~\cite{kiselev}.}
\label{fig:isoc}
\end{figure}

We now briefly consider some thermodynamic properties of the SW fluid. 
Figure~\ref{fig:zcomp} displays the compressibility factor $\beta P/\rho$ and 
the internal energy per particle $U^{*}$ for $\delta=0.1$ along the isotherms
$T^{*}=2$ and $T^{*}=1$, compared with simulation results~\cite{bergenholtz}. 
The agreement is very satisfactory both at low and high density, although it 
is fair to point out that the temperatures at which the simulations were 
performed can be considered quite high, since even $T^{*}=1$ is still more than
twice the expected critical value reported in Tab.~\ref{tab:crit}. 
A comparison encompassing temperatures closer to $T_{c}$ is shown in 
Fig.~\ref{fig:isoc}, where $\beta P/\rho$ and $U^{*}$ for $\delta=0.25$ are 
shown along the isochore $\rho^{*}=0.42$ as a function of the inverse 
temperature $\beta$, and compared with simulation results~\cite{kiselev}.
The agreement is again quite good down to the lowest temperature attainable by 
the SCOZA, i.e., that at which the isochore hits the high-density branch of the
spinodal curve. Inspection of Fig.~2 of~\cite{kiselev} suggests that the 
simulation data at the lowest temperatures, roughly for 
$\beta\gtrsim 1.33$, would correspond to unstable states in the thermodynamic
limit.    

\begin{figure}
\includegraphics[width=10cm]{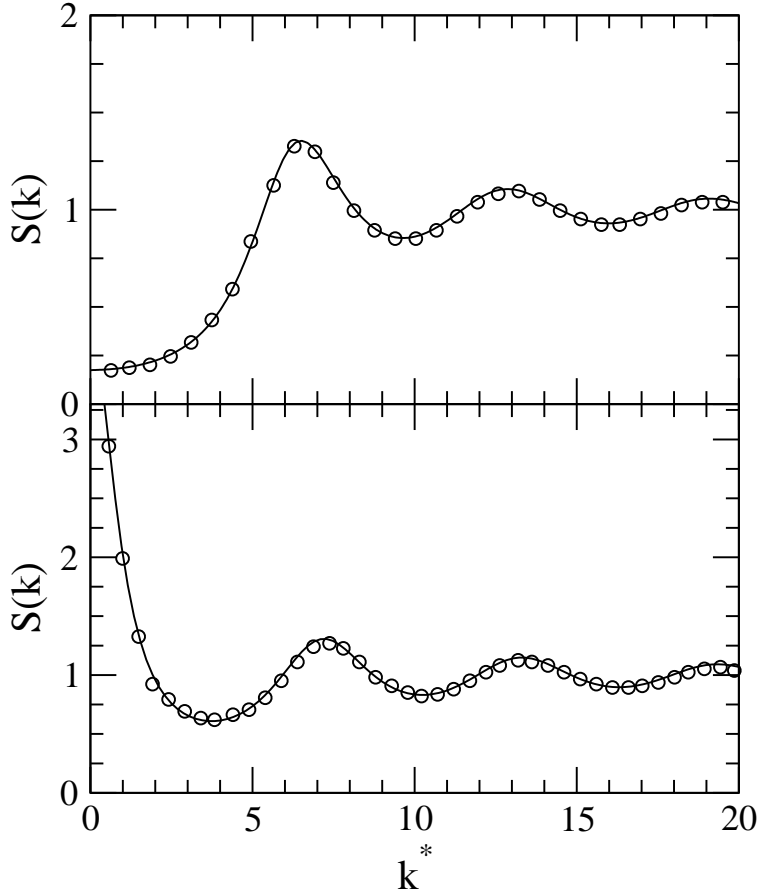}
\caption{Structure factor of the SW fluid with $\delta=0.1$ for 
$T^{*}=2$, $\rho^{*}=0.5$ (upper panel) and $T^{*}=0.5$, $\rho^{*}=0.229$ 
(lower panel). Solid line: SCOZA. Circles: simulation data by Shukla 
et al.~\cite{shukla} extracted from their Figs.~6 and 8.}
\label{fig:stru}
\end{figure}

\begin{figure}
\includegraphics[width=10cm]{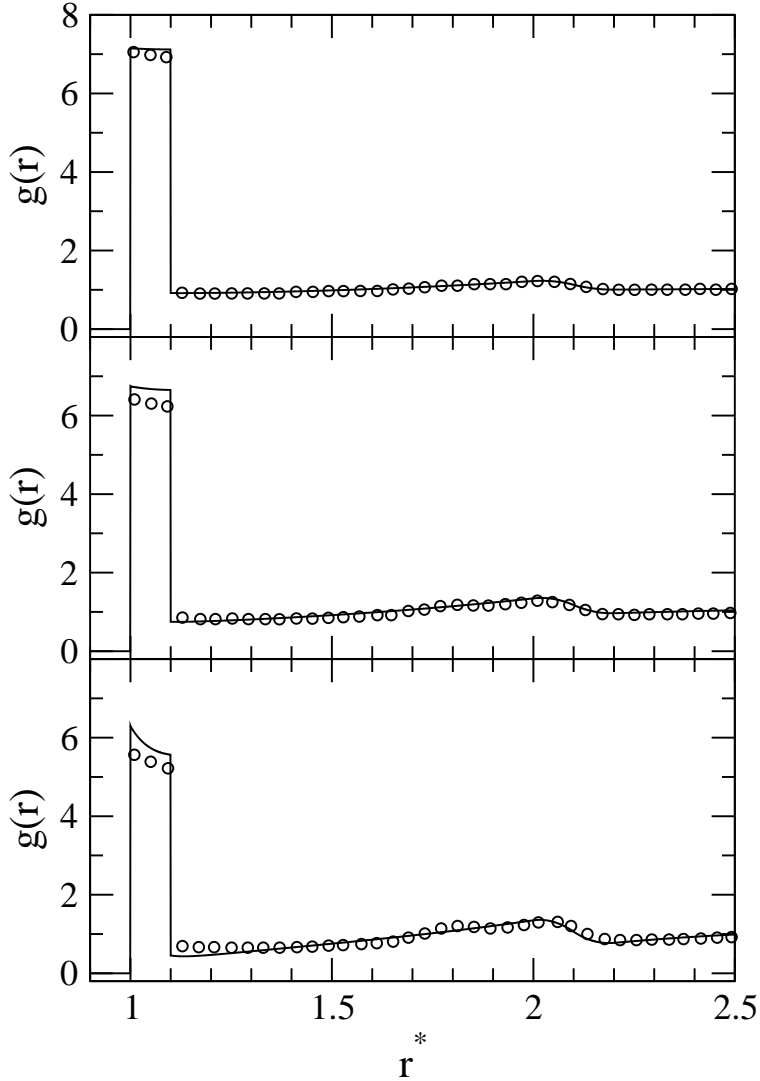}
\caption{Radial distribution function of the SW fluid with 
$\delta=0.1$ for $T^{*}=0.5$ and $\rho^{*}=0.2$ (top panel), $\rho^{*}=0.4$ 
(central panel), $\rho^{*}=0.8$ (bottom panel). Solid line: SCOZA. Circles:
simulation data by Largo et al.~\cite{santos} extracted from their Fig.~6.}
\label{fig:gcorr}
\end{figure}

It is also worthwhile examining how the present theory performs for the 
correlations of the fluid. Figure~\ref{fig:stru} shows the structure factor
$S(k)$ for a SW with $\delta=0.1$ and two different states, one at  
moderate density $\rho^{*}=0.5$ and relatively high temperature $T^{*}=2$, and 
another at lower density $\rho^{*}=0.229$ and temperature $T^{*}=0.5$ near
the critical value. In both cases, the SCOZA reproduces quite accurately the
simulation results~\cite{shukla}. 
The correlations in real space are considered in Fig.~\ref{fig:gcorr}, where
the radial distribution function $g(r)$ predicted by the SCOZA, again for 
$\delta=0.1$ and $T^{*}=0.5$, is compared with simulation data~\cite{santos} 
at the densities $\rho^{*}=0.2$, $\rho^{*}=0.4$, $\rho^{*}=0.8$. 
At low density, the theory 
agrees closely with the simulation, both inside and outside the well region. 
This is not surprising since, as pointed out in Sec.~\ref{sec:intro}, the SCOZA
for the SW potential becomes exact at low density, while this is not
the case for other tail interactions. On the other hand, as $\rho$ is increased 
the SCOZA overestimates both the contact value $g(\sigma^{+})$ of $g(r)$, and
the amplitude of the discontinuity at the well boundary 
$\xi\equiv (1+\delta)\sigma$. This also testifies the quite special status 
of the SW potential within the SCOZA: in fact, for  
the HCY potential~\cite{scozayuk4}  
the SCOZA was found to underestimate $g(\sigma^{+})$, as one would expect from a theory whose $c(r)$ is linear in the perturbation potential. In the present 
case, the behavior at high density is instead reminiscent of that of the
``non-linear ORPA'' studied in~\cite{orpanl}. Another feature worth attention
is the hump displayed by the simulation $g(r)$ for $\rho^{*}=0.8$ and 
$r^{*}\simeq 1.8$, which is the precursor of the discontinuity for 
$r^{*}=\sqrt{3}$ displayed by $g(r)$ in the sticky 
limit~\cite{miller2,malijevsky}. This is not reproduced by the SCOZA, most 
likely because of the assumption implied by Eq.~(\ref{closure}) that $c(r)$ has 
the same range as the potential. 
\begin{figure}
\includegraphics[width=10cm]{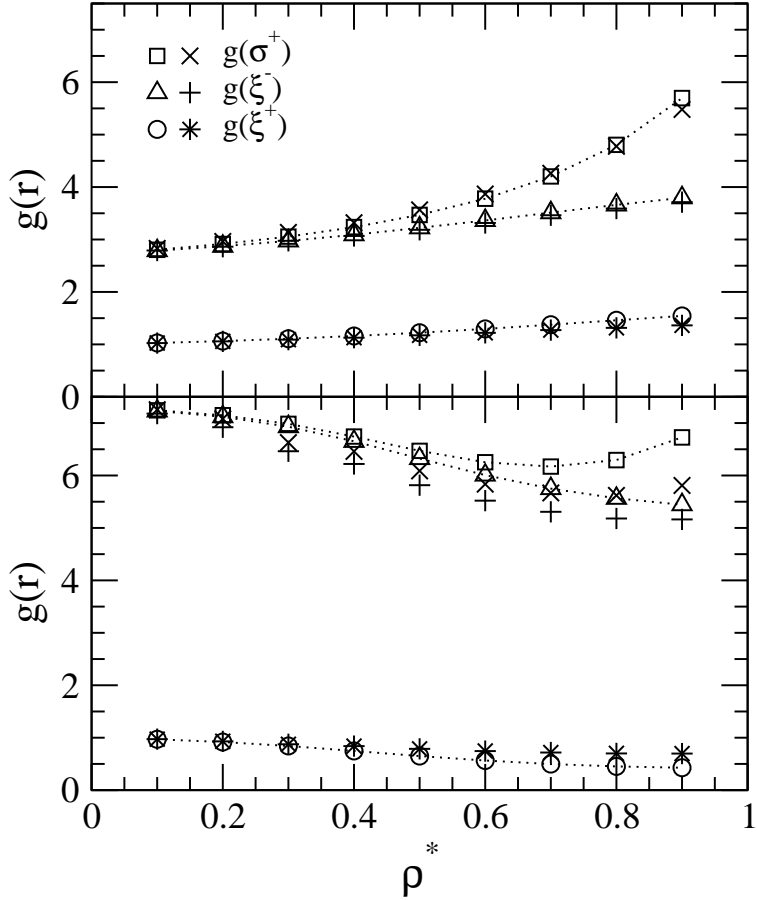}
\caption{Radial distribution function of the SW fluid with 
$\delta=0.1$ for $r\to\sigma^{+}$, 
$r\to\xi^{-}$, $r\to\xi^{+}$, where $\xi\equiv (1+\delta)\sigma$ 
is the value of $r$ beyond which the potential vanishes. Squares, triangles, 
and circles represent the SCOZA results for $g(\sigma^{+})$, $g(\xi^{-})$,
and $g(\xi^{+})$ respectively. Crosses, pluses, and starbursts represent
the simulation data obtained by Largo et al.~\cite{santos} for the same 
quantities. Dotted lines are a guide to the eye. Upper panel: $T^{*}=1$ 
isotherm. Lower panel: $T^{*}=0.5$ isotherm.}   
\label{fig:contact}
\end{figure}

A closer look at the $g(r)$ at contact and at the well boundary is given in 
Fig.~\ref{fig:contact}, where $g(\sigma^{+})$, $g(\xi^{-})$, $g(\xi^{+})$ for
$\delta=0.1$ are shown along the isotherms $T^{*}=0.5$ and $T^{*}=1$.  
As one may expect, the aforementioned discrepancy with the simulation data that
sets out at large $\rho$ becomes less severe as $T$ is increased, the results
for $T^{*}=1$ being in good agreement with the simulation~\cite{santos} even
at the highest $\rho$ considered.    

\section{The sticky limit of the SCOZA}
\label{sec:sticky}

\begin{figure}
\includegraphics[width=10cm]{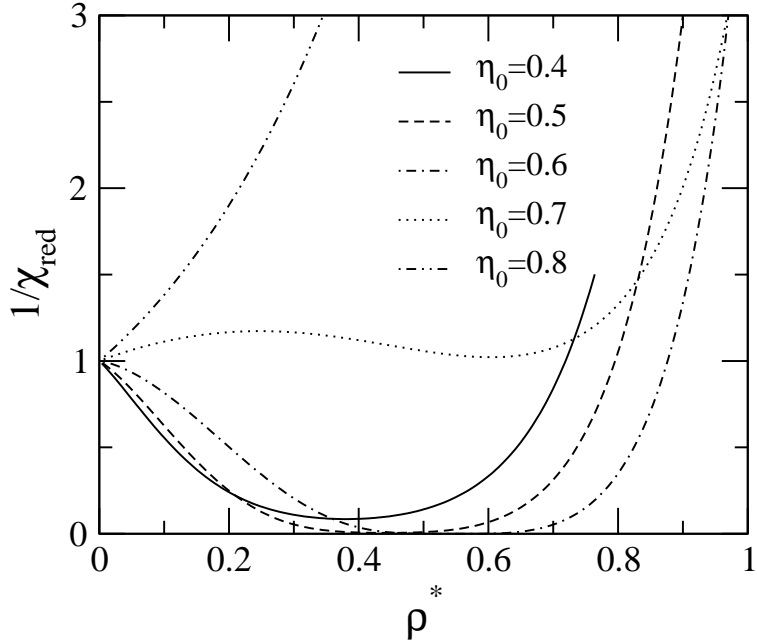}
\caption{Inverse reduced compressibility $1/\chi_{\rm red}$ of the SHS fluid
as a function of the reduced density $\rho^{*}$ as found via the SCOZA 
equation~(\ref{scozasticky}) with a high-density boundary condition computed
in the PY approximation~\cite{baxter}. The different lines refer to the 
boundary packing fractions $\eta_{0}=\pi\rho_{0}^{*}/6$ given in the legend.
The stickiness parameter is set at $\tau=0.11$, which is close to the critical 
value according to numerical simulations~\cite{miller}.} 
\label{fig:chisticky}
\end{figure}

As discussed in Sec.~\ref{sec:results}, the numerical solution of the SCOZA for
the SW potential developed here becomes unpredictive for wells
such that $\delta\leq 0.05$, still considerably wider than the HCY 
potentials that were considered in~\cite{foffi,scozaorea}. However, in the 
limit $\delta\to 0$ it is possible to study the behavior of the SCOZA by
adopting a different standpoint, namely, by imposing thermodynamic consistency
on the solution to Eqs.~(\ref{oz}), (\ref{closure}) {\em after} the limit
$\delta\to 0$ has been taken. The clue to this procedure lies in the 
observation made in~\cite{gazzillo} that in the limit $\delta\to 0$, any OZ 
closure like Eq.~(\ref{closure}), such that one has $h(r)=-1$ for $r<\sigma$ 
and $c(r)=0$ for $r>(1+\delta)\sigma$, gives a fixed form for $c(r)$: 
\begin{equation}
c(r)=\left\{-\, \frac{\eta\lambda}{r^{*}}-a^{2}+\eta\left[6(a+b)^{2}-a\lambda
\right]r^{*}
-\frac{\eta a^{2}}{2}{r^{*}}^{3}\right\}\Theta(1-r^{*})+\frac{\lambda}{12} \, 
\delta(r^{*}-1)   \, ,
\label{csticky}
\end{equation}
where $\Theta(x)$ and $\delta(x)$ are respectively the Heaviside and Dirac 
functions of argument $x$, and we have set 
$\eta\equiv \pi\rho\sigma^{3}/6$. The quantities $a$ and $b$ are given by
\begin{eqnarray}
a & = & \frac{1+2\eta}{(1-\eta)^{2}}-\frac{\lambda\eta}{1-\eta} \, ,
\label{asticky} \\
b & = & -\frac{3\eta}{2(1-\eta)^{2}}+\frac{\lambda\eta}{2(1-\eta)} \, .
\label{bsticky}
\end{eqnarray} 
Therefore, in the limit $\delta\to 0$ all the specific features of the OZ 
closure considered are conveyed into a single, state-dependent parameter 
$\lambda$. The latter is related to the stickiness parameter $\tau$ defined in
Eq.~(\ref{tau}) and to the contact value of the cavity function 
$y(r)=g(r)\exp{[\beta v(r)]}$ by the expression
\begin{equation} 
\lambda=\frac{y({\sigma})}{\tau} \, .
\label{lambda}
\end{equation}
By applying the compressibility rule~(\ref{comp}) to Eq.~(\ref{csticky}), one
obtains
\begin{equation}
\frac{1}{\, \chi_{\rm red}}=a^{2} \, ,
\label{compsticky}
\end{equation}
whereas the energy route gives~\cite{watts}
\begin{equation}
\frac{\beta A}{N} = \frac{\beta A_{\rm HS}}{N} 
-\eta\int_{\tau}^{+\infty}\!\!d\tau^{\prime}\, 
\frac{\lambda(\eta,\tau^{\prime})}{\tau^{\prime}} \, .
\label{energysticky}
\end{equation}
If the form of $\lambda(\eta, \tau)$ obtained by the PY closure
is substituted into Eqs.~(\ref{csticky})-(\ref{energysticky}),   
the aforementioned solution of the SHS model due to Baxter~\cite{baxter} is 
recovered. On the other hand, as discussed in detail in~\cite{gazzillo}, 
different choices for $\lambda$ are also possible. Since one of the most 
problematic features of the original SHS model is its lack of thermodynamic 
consistency, it is tempting to determine $\lambda$ via the SCOZA, thus 
enforcing consistency between the compressibility and energy routes given by 
Eqs.~(\ref{compsticky}), (\ref{energysticky}). If we introduce the quantities
\begin{eqnarray}
\Lambda & = & \eta^{2}\lambda \, ,
\label{lambdacap} \\
t & = & \frac{1}{\tau} \, , 
\label{taurec}
\end{eqnarray}
the SCOZA equation reads
\begin{equation}
{\cal D}(\eta, t)\, t \frac{\partial\Lambda}{\partial t}=
\frac{\partial^{2}\!\Lambda}{\partial\eta^{2}} \, ,
\label{scozasticky}
\end{equation}
where ${\cal D}(\eta, t)$ is given by
\begin{equation}
{\cal D}(\eta, t)=2\, \frac{\eta\, (1+2\eta)-(1-\eta)\Lambda}
{[\, \eta\, (1-\eta)]^{3}} \, .
\label{diffsticky}
\end{equation}
The PDE~(\ref{scozasticky}) can be solved numerically by the finite-difference
scheme described in the Appendix of~\cite{scozaising}, or the similar one 
given here by Eq.~(\ref{pdediscr}); unlike in the case of non vanishing 
$\delta$,
the diffusion coefficient is given explicitly by Eq.~(\ref{diffsticky}), so 
there is no need to evaluate it by finite differences as 
in Eq.~(\ref{diffdiscr}). As usual, for the problem to be completely specified,
Eq.~(\ref{scozasticky}) must be complemented with the initial and boundary 
conditions. As for the initial condition, we observe that the numerical
integration cannot be started exactly from $t=0$, because for this value of
$t$ the diffusion coefficient $1/(t{\cal D})$ would become infinite. Instead,
the integration was started from a small, but non null $t_{0}>0$, using 
as the initial condition for $\Lambda$ the leading term in an expansion in 
powers of $t_{0}$
\begin{equation}
\Lambda(\eta, t_{0})\simeq t_{0}\, \varphi(\eta) \, .
\label{lambdasmall} 
\end{equation}    
By substituting Eq.~(\ref{lambdasmall}) into Eq.~(\ref{scozasticky}) we find
that the $\eta$-dependent term $\varphi(\eta)$ must satisfy the ordinary 
differential equation~\footnote{Although for $\tau\to \infty$ the contact value
of the radial distribution function coincides with the PY result, $y(\sigma)$
does not because of the double limiting procedure $\delta\to 0$, 
$\tau\to\infty$ in the prefactor $e^{\beta v(\sigma^{+})}$.} 
\begin{equation}
\frac{d^{2}\varphi}{d\eta^{2}}=2\, \frac{1+2\eta}{\eta^{2}(1-\eta)^{3}}\, 
\varphi \, .
\label{scozasmall}
\end{equation}
This is solved numerically with the conditions
\begin{eqnarray}
& & \varphi(0)=\left.\frac{d\varphi}{d\eta}\right|_{\eta=0}\!\!=0 \, ,
\label{init1} \\
& & \left.\frac{d^{2}\varphi}{d\eta^{2}}\right|_{\eta=0}\!\!=2 \, , 
\label{init2}
\end{eqnarray}
that follow from Eqs.~(\ref{lambda}), (\ref{lambdacap}), (\ref{lambdasmall})
and the property~\cite{gazzillo} 
\begin{equation}
\lim_{\eta\to 0} y(\sigma)=1 \, .
\label{ysmall}
\end{equation}
We have verified {\em a posteriori} that the numerical solution 
of Eq.~(\ref{scozasticky}) is virtually independent of the particular 
choice of $t_{0}$, as long as the latter is sufficiently small. 
Equations~(\ref{lambda}), (\ref{lambdacap}), also imply for the low-density 
boundary condition
\begin{equation}
\Lambda(\eta=0, t)=0 \mbox{\hspace{2cm} for every $t$.}
\label{lambdalow}
\end{equation}
As observed in Sec.~\ref{sec:theory}, the choice of the high-density boundary 
condition entails a certain amount of arbitrariness. A natural 
possibility is using for $\lambda$ the expression $\lambda_{\rm Bax}$ obtained 
by the analytical solution of the SHS model within the PY 
approximation~\cite{baxter}:
\begin{equation}
\Lambda(\eta_{0}, t)=\eta_{0}^{2}\lambda_{\rm Bax}(\eta_{0}, t) 
\mbox{\hspace{2cm} for every $t$},  
\label{lambdahigh}  
\end{equation}
where $\eta_{0}$ denotes the position of the high-density boundary.
Quite disappointingly, although perhaps not surprisingly, the outcome of the 
numerical integration of Eq.~(\ref{scozasticky}) with 
the initial and boundary conditions given by Eqs.~(\ref{lambdasmall}), 
(\ref{lambdalow}), (\ref{lambdahigh}) suffers from the same problem experienced
in Sec.~\ref{sec:results} for wells of non vanishing width, namely, a strong
dependence of the results on the position of $\eta_{0}$. This is illustrated
in Fig.~\ref{fig:chisticky}, where the inverse reduced compressibility 
$1/\chi_{\rm red}$ obtained for different choices of $\eta_{0}$ has been
plotted along the isotherm $\tau=0.11$, which is close to the critical one
according to numerical simulations~\cite{miller}. For the solution 
to be deemed reliable, the various curves, at least for $\eta_{0}$ greater than
a certain threshold, ought to lie near one another --- a requirement that is
manifestly {\em not} satisfied. Some of the isotherms are not even consistent 
with $\tau=0.11$ being close to the expected critical value. We have also 
verified that this state of affairs is not changed by different choices of 
the high-density boundary condition.

\begin{figure}
\includegraphics[width=10cm]{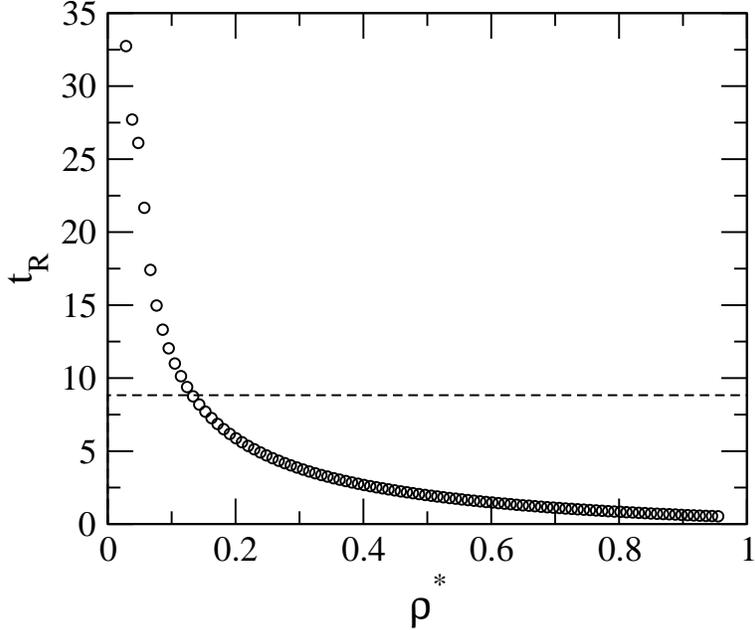}
\caption{Radius of convergence $t_{\rm R}$ of the power series~(\ref{series}), 
estimated by means of the ratio convergence test, as a function of the reduced 
density $\rho^{*}$. The horizontal dashed line marks the value of the inverse
stickiness $t$ corresponding to the critical point according to numerical 
simulations~\cite{miller}.}
\label{fig:radius}
\end{figure}

\begin{figure}
\includegraphics[width=10cm]{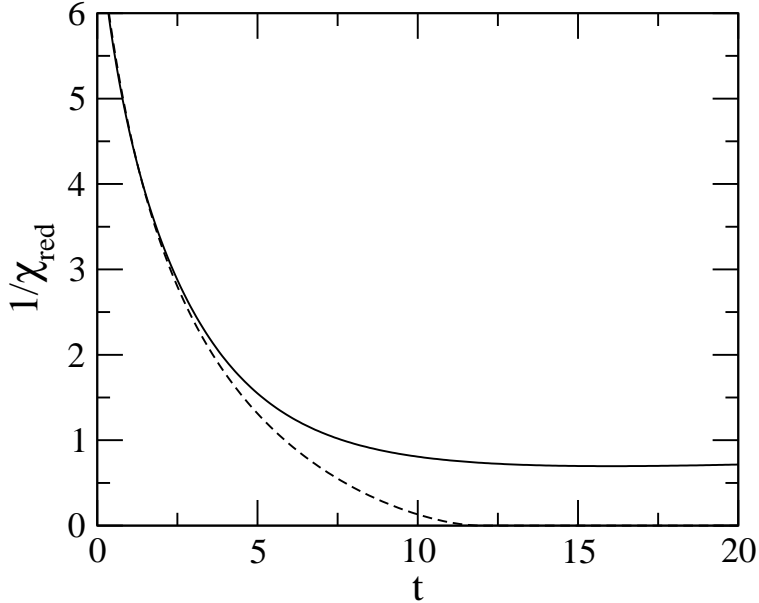}
\caption{Inverse reduced compressibility $1/\chi_{\rm red}$ of the SHS fluid
as a function of the inverse stickiness $t$ along the $\eta=0.25$ 
isochore. The compressibility route of the PY approximation (dashed
line) predicts the compressibility to diverge at $t\simeq 12$. According to the
solution of Eq.~(\ref{scozasticky}) evaluated by means of the Borel summation
method (solid line), the isochore does not cross the spinodal at all.}
\label{fig:borel}
\end{figure}

We can, however, resort to a different method to solve Eq.~(\ref{scozasticky}),
that does not require the high-density boundary to be specified. 
This is made possible by the fact that, unlike in the case of wells of 
non vanishing width discussed in Secs.~\ref{sec:theory}, \ref{sec:results}, 
the diffusion coefficient of 
Eq.~(\ref{scozasticky}) is known explicitly as a function of $\eta$ and $\tau$.
Let us assume that $\Lambda$ can be expressed as a power series in $t$
\begin{equation}
\Lambda(\eta, t)=\sum_{n=1}^{\infty}t^{n}\omega_{n}(\eta) \, ,
\label{series}
\end{equation}
where the zeroth-order term has been omitted, since $\Lambda$ must vanish for
$t\to 0$. By inserting Eq.~(\ref{series}) into Eq.~(\ref{scozasticky}) and
equating the terms of the same degree in $t$, Eq.~(\ref{scozasticky}) is 
replaced by an infinite set of ordinary differential equations:
\begin{equation}
\frac{d^{2}\omega_{n}}{d\eta^{2}}=2n\frac{1+2\eta}{\eta^{2}(1-\eta)^{3}} \, 
\omega_{n}-\frac{2}{\eta^{3}(1-\eta)^{2}}\sum_{j=1}^{n-1}
j\, \omega_{j}\omega_{n-j} \, ,
\label{ode}
\end{equation}
which gives back Eq.~(\ref{scozasmall}) for $n=1$. Since this set  
contains only backward references, it can be integrated numerically to any 
desired order $\bar{n}$. Only the low-density behavior of $\omega_{n}$
is required.
Specifically, one needs the values of $\omega_{n}$, $d\omega_{n}/d\eta$, and
$d^{2}\omega_{n}/d\eta^{2}$ at $\eta=0$ for $n\leq \bar{n}$. If it is assumed 
that $\omega_{n}$ can be expanded in power series of $\eta$, it is found that 
for $n>6$ all the above quantities are in fact vanishing, while for $n\leq 6$ 
they can be obtained from exact diagrammatic expansions available in 
the literature~\cite{gazzillo,post}. Unfortunately, the radius of convergence
$t_{\rm R}$ of the series~(\ref{series}) corresponding to most values of the 
packing fraction $\eta$ comes out to be too small for it to be of direct 
usefulness. 
This is apparent from Fig.~\ref{fig:radius}, which shows how 
$t_{\rm R}$ rapidly decreases as the density is increased. In particular, 
at the critical value of $t$ predicted by simulations~\cite{miller}, $\rho^{*}$ 
should be well below $0.2$ to obtain convergence. 
This is again disappointing, yet not
uncommon: a power series obtained as a formal solution of a differential 
equation may easily turn out to be divergent. In those cases, one can try a 
resummation technique in order to provide an analytic meaning to such a 
divergent series. To this end, we have employed Borel summation 
method~\cite{kawai}, whereby one starts from the exact equality
\begin{equation}
\Lambda(\eta, t)=\sum_{n=1}^{\infty}n! \frac{t^{n}}{n!}\, \omega_{n}(\eta)=
\frac{1}{t}\sum_{n=1}^{\infty}\int_{0}^{+\infty}\!\!ds\, e^{-s/t} \, 
\frac{{s}^{n}}{n!} \, \omega_{n}(\eta) 
\label{borel}
\end{equation}
and formally interchanges the series and the integral operators to obtain
\begin{equation}
\Lambda_{\rm B}(\eta, t)=\frac{1}{t}\int_{0}^{+\infty}\!\! ds\, e^{-s/t} \, 
\tilde{\Lambda}(\eta, s) \, ,
\label{lambdaborel}
\end{equation}
where $\tilde{\Lambda}(\eta, s)$ is the regularized power series
\begin{equation}
\tilde{\Lambda}(\eta, s)=\sum_{n=1}^{\infty}\frac{s^{n}}{n!}\, \omega_{n}(\eta) 
\, ,
\label{lambdatilde}
\end{equation}
that has an improved radius of convergence because of the factorial in the 
denominator. Where the original series is convergent, the two functions 
$\Lambda$ and $\Lambda_{\rm B}$ are in fact identical. The latter, however, 
yields a finite value in a much larger domain, providing the analytic 
continuation of the function defined by the original series. 
If one is willing to trust the Borel sum $\Lambda_{\rm B}$ in this enlarged
region, then a puzzling result is found: the liquid-vapor phase transition 
disappears, at least for reasonable values of the packing fraction and 
temperature. In Fig.~\ref{fig:borel} the inverse reduced compressibility of 
the fluid $1/\chi_{\rm red}$ is shown as a function of $t$ along the 
$\eta=0.25$ isochore. 
While the compressibility route of the PY 
approximation predicts the compressibility to diverge around 
$\tau\simeq 0.08$, the solution of Eq.~(\ref{scozasticky}) evaluated
by means of Borel summation method gives instead a finite $\chi_{\rm red}$ even
if the temperature is lowered further. This result is qualitatively similar to
that found by direct numerical integration of Eq.~(\ref{scozasticky}) when 
the high-density boundary $\eta_{0}$ is pushed to (unphysically) high values, 
but is clearly at odds not only with the PY solution of the SHS 
model~\cite{baxter}, but also with the simulation results~\cite{miller}. 
The present
investigation supports the conclusion that in the limit of vanishing well 
width $\delta\to 0$, the SCOZA critical temperature decreases too rapidly to 
give a finite value of $B_{2}(T_{c})$, so that $B_{2}(T)$ or, equivalently,  
the stickiness $\tau$, do not represent a convenient way to measure the 
temperature in this limit. Imposing thermodynamic consistency on the SHS model
does not lead to a better quantitative agreement with simulation compared 
to the highly inconsistent PY solution, but merely makes the 
liquid-vapor transition disappear.       

\section{Conclusions}
\label{sec:conclusions}

We have performed an investigation of the SCOZA for SW potentials, 
focused on the regime of narrow wells which is the most relevant for the 
modelization of colloidal particles. This study was prompted by the lack of a 
theory able to account quantitatively for the liquid-vapor phase transition
in a system of sticky hard spheres: in the well known and widespreadily used 
PY approximation~\cite{baxter}, the various routes to the 
thermodynamics of the fluid are inconsistent with one another, and agree only
qualitatively with the results of computer simulation~\cite{miller}. On the
other hand, for HCY potentials the SCOZA can be solved 
semi-analytically, and has already been usefully applied to narrow-range 
tails~\cite{foffi,scozaorea}. 
The rationale for
turning to the SW potential within the SCOZA is that this interaction
has the unique feature of having the SCOZA become exact at low density, thereby 
recovering the correct $B_{2}(T)$ which plays such an important role for narrow
interactions. 

However, the results presented here show that this peculiarity turns more into 
a liability than an asset. For wells of non vanishing width $\delta$, after 
extending the study already made in~\cite{paschinger} to (slightly) smaller 
$\delta$, we cannot but confirm the conclusions of that work, namely: i) the 
results are found to be very sensitive to the boundary condition at high 
density that is needed to integrate the SCOZA PDE, actually much more so than 
in the case of HCY potentials of comparable range. Such an undesired feature 
can be tamed by 
pushing the density of the boundary to very high values, but this has the 
effect of making the numerical solution of the PDE unfeasible for 
$\delta\lesssim 0.05$. ii) In the range of $\delta$ for which meaningful 
results can be extracted, the overall agreement with simulation data, at least
for the phase diagram, is actually worse than what was found for the HCY 
potential~\cite{scozayuk4,foffi,scozaorea}, although a sharp assessment of the 
performance of the SCOZA is made
difficult by the considerable discrepancies between different simulations.  

If the sticky limit is taken from the outset, and subsequently thermodynamic
consistency is imposed, the SCOZA PDE is obtained explicitly in closed form,
but the high-density boundary problem is again found to be looming. If the
PDE is turned into an infinite set of ordinary differential equations and the
low-density behavior alone is imposed, the liquid-vapor phase boundary 
disappears altogether. This supports the conclusion 
that as the attractive potential range vanishes, the SCOZA 
critical 
temperature will decrease more rapidly than what is necessary to yield a
finite $B_{2}(T_{c})$. 

Stating that the SCOZA is inadequate for SW interactions in the 
sticky limit is merely an 
observation, and does not shed any light on the reason behind the fact. In this
respect, one should recall the result of Stell~\cite{stell} mentioned in 
Sec.~\ref{sec:intro}, who showed that 
the SHS model is strictly speaking ill-defined if a class of
problematic configurations of particles is not ruled out, for instance by 
introducing a small amount of polidispersity into the system. Can it be that
the (possibly more accurate) SCOZA approach exposes the malicious effect of the
diverging clusters, which are simply neglected in the PY 
approximation? We regard this as unlikely, since the two approximations are 
quite similar in their handling of the correlation functions of the fluid, 
giving the same functional form for $c(r)$ in the sticky 
limit~\cite{gazzillo}. What we deem most probable, instead, is that the failure
is a combined effect of the approximate closure~(\ref{closure}) and the 
consistency requirement~(\ref{consist}). Indeed, the closure is likely to miss
a number of singular features, i.e., discontinuities and Dirac-$\delta$ peaks, 
that show up in the real correlation 
function~\cite{miller2,malijevsky}: the isothermal compressibility, being 
dependent on the volume integral of the same function, suffers from the same
defect. On the contrary, the internal energy of the fluid depends on the 
contact value of the cavity function alone: when thermodynamic consistency is 
imposed,
the latter has to compensate for the missing contributions, unnaturally pushing
away the liquid-vapor phase boundary. We stress that at the moment this is a 
mere, untested hypothesis.    

One may hope that the strong dependence of the results  
on the high-density boundary condition found for narrow SW 
potentials will become less drastic turning to a different form of the 
interaction. In fact, as observed above, for the HCY potential 
this unwelcome feature, albeit still there, is nevertheless much less 
pronounced. If this is actually the case, then the fully numerical solution of 
the SCOZA which has been developed here and the one previously developed 
in~\cite{paschinger}, could still be usefully applied to different forms 
of short-range potentials, not to mention the instance of interactions that are 
not of short range~\cite{paschinger08}. Moreover, this solution could be 
helpful in the context of the renormalization-group based hierarchical 
reference theory (HRT)~\cite{hrt}, specifically for enforcing the core 
condition on $g(r)$ exactly (of course within the limit of numerical accuracy).
In fact, in the smooth cut-off formulation of the HRT~\cite{smooth}, similarly 
to the SCOZA, this condition can be (and has been) implemented analytically 
for a HCY potential, but for a generic tail potential a numerical solution 
would again be needed. Also, this is always the case for the sharp-cut off HRT,
irrespective of the form of the tail potential. In the latter theory, the core
condition was implemented by introducing a number of 
approximations~\cite{giri}, which however affect the accuracy of this approach 
for short-range interactions~\cite{scozayuk4}. A better procedure such as that
developed here would certainly be more adequate.       

As for the sticky limit, the results presented in this work suggest that, if 
thermodynamic consistency is to be employed as a way to
improve upon the accuracy of the PY solution for the SHS potential~(\ref{shs}),
one should first get rid of the approximation common to the SCOZA and the PY 
equation: namely, that the
contribution to the direct correlation function $c(r)$ due to the tail 
interaction $w(r)$ has the same range as $w(r)$ itself, and therefore vanishes 
wherever $w(r)$ does. In this respect, it is worthwhile recalling that the 
same approximation is used in conjunction with thermodynamic consistency not 
only in the SCOZA, but also in the HRT~\cite{smooth}. Therefore, if the 
combination of these ingredients is bound to become dangerous for vanishing 
attraction range, it is likely that this would affect the HRT as well, and that
both the SCOZA and the HRT would benefit from a solution to this problem.  

\begin{acknowledgments}
We would like to dedicate this paper to Bob Evans on his $65^{\rm th}$ 
birthday. We thank
Pedro Orea for sending us simulation data for the phase diagram of the
SW fluid and the Italian Ministry of Education and Research (MIUR) for PRIN 
2008 funding. D.~P. would also like to thank Paolo Giaquinta for stimulating 
the investigation of the behavior of the SCOZA for narrow attractive 
interactions and for his unflagging interest in this problem. 
\end{acknowledgments}

\appendix

\section{Finite-difference discretization of the SCOZA PDE}
\label{sec:appint} 

Equation~(\ref{pde}) is a nonlinear, diffusive-like PDE, where the quantity 
$D(\rho, u)$ given by Eq.~(\ref{diff}) plays the role of the reciprocal of the
diffusion coefficient. This PDE is discretized in the following way:
\begin{equation}
D(\rho_{j}, u_{j}^{n+1/2})\, \frac{u_{j}^{n+1}-u_{j}^{n}}{\Delta\beta} = 
\frac{\rho_{j}}{2}\left[\frac{u_{j+1}^{n+1}-2u_{j}^{n+1}+u_{j-1}^{n+1}}
{(\Delta\rho)^{2}} + \frac{u_{j+1}^{n}-2u_{j}^{n}+u_{j-1}^{n}}
{(\Delta\rho)^{2}}\right]
\label{pdediscr}
\end{equation}
and
\begin{equation}
D(\rho_{j}, u_{j}^{n+1/2}) = \frac{\left(\displaystyle{\frac{1}
{\, \chi_{\rm red}}}\right)_{\! j}^{\! n+1} - \left(\displaystyle{\frac{1}
{\, \chi_{\rm red}}}\right)_{\! j}^{\! n}}{u_{j}^{n+1}-u_{j}^{n}} \, ,
\label{diffdiscr}
\end{equation}
where subscripts refer to the density $\rho$ and superscripts to the inverse 
temperature $\beta$ in the $(\rho, \beta)$ grid, so for instance
$u_{j}^{n}\equiv u(\rho_{j},\beta_{n})$,
and $\Delta\beta$, $\Delta\rho$ are respectively the inverse temperature
and density steps used in the discretization. 

Equation~(\ref{pdediscr}) has been 
solved by an implicit algorithm as follows: assuming that $u_{j}^{n}$ is known
at a certain inverse temperature $\beta_{n}$ for every $\rho_{j}$, a guess for 
$u_{j}^{n+1}$ at the ``new'' temperature $\beta_{n+1}$ is given. This is 
determined by solving Eqs.~(\ref{oz}), (\ref{closure}) with the approximation
$K_{n+1}\simeq K_{n}+\Delta\beta$. The values of the inverse reduced 
compressibility $(1/\chi_{\rm red})_{j}^{n+1}$ corresponding to $u_{j}^{n+1}$
are then obtained by solving Eqs.~(\ref{oz}), (\ref{closure}) at fixed density
$\rho$ and internal energy $u$ via the method described in 
Sec.~\ref{sec:theory} and Appendix~\ref{sec:appopt}.
These values are used in Eq.~(\ref{diffdiscr}) to obtain a trial value of
$D(\rho_{j}, u_{j}^{n+1/2})$, which is inserted into Eq.~(\ref{pdediscr}) to 
give a closed linear system of equations for $u_{j}^{n+1}$ at $\beta_{n+1}$. 
The values of $u_{j}^{n+1}$ thus obtained replace the initial guess, and the 
procedure is iterated until convergence is obtained. The $u_{j}^{n+1}$ at 
convergence are fed back into the algorithm to obtain $u_{j}^{n+2}$, and so on.

\section{Numerical optimization of the free energy functional}
\label{sec:appopt}

The most straightforward method to minimize the functional ${\cal S}$ defined 
by Eq.~(\ref{s}) is the steepest
descent~\cite{pastore}, in which $\phi(r)$ and $K$ are updated iteratively 
by moving downhill
in the direction opposite to that of the gradient of ${\cal S}$, namely
\begin{eqnarray}
\phi_{l+1}(r) & = & \phi_{l}(r) - \lambda \, \Delta h_{l}(r) 
\mbox{\hspace{1.5cm}} r<\sigma \, ,
\label{steep1} \\
K^{*}_{l+1} & = & K^{*}_{l} - \lambda \, \Delta V_{l} \, .
\label{steep2}
\end{eqnarray}
Here the indexes $l$, $l+1$ refer to the values of $\phi(r)$ and $K$ at two 
consecutive iterations, $\lambda$ (not to be confused with 
the state-dependent 
function $\lambda(\eta, \tau)$ introduced in Sec.~\ref{sec:sticky}) 
is a parameter determining the step size, and we have set 
\begin{equation}
\Delta V = -\rho\!\int\!\!d^{3}{\bf r} \, \Delta h(r) w^{*}(r) + \Delta U^{*} 
\, . 
\label{deltav}
\end{equation}
The asterisks in Eqs.~(\ref{steep2}), (\ref{deltav}) denote reduced quantities
$K^{*}=K\epsilon$, $\Delta U^{*}=\Delta U/\epsilon$, $w^{*}(r)=w(r)/\epsilon$, 
where $\epsilon$ is the energy scale of $w(r)$. For the SW potential
considered here, $\epsilon$ coincides with the well depth. Note that, for 
$\lambda$ to be the same dimensionless quantity in Eqs.~(\ref{steep1}) and 
(\ref{steep2}), the differentiation of ${\cal S}$ has to be performed with 
respect to $K^{*}$ and $\sqrt{\rho}\phi(r)$. 

To implement the algorithm, one starts from a guess for $K$ and $\phi(r)$ for 
$r<\sigma$, and determines $\phi(r)$ for $r>\sigma$ by the relation 
$\phi(r)=-K w(r)$. 
In the numerical solution of the SCOZA, the guess at the inverse 
temperature $\beta_{n+1}$ and density $\rho_{j}$ was given by 
$\phi(r, \rho_{j}, \beta_{n+1})\simeq \phi(r, \rho_{j}, \beta_{n})$, 
$K(\rho_{j}, \beta_{n+1})\simeq K(\rho_{j}, \beta_{n})+\Delta\beta$. 
By switching to reciprocal space, the Fourier transform 
of $\Delta h(r)$ is determined via the OZ equation as
\begin{equation}
\Delta \hat{h}(k) = S_{\rm HS}(k)S(k)\hat{\phi}(k) \, ,
\label{deltah}
\end{equation}
where $S(k)$ is given by Eq.~(\ref{srpa}). Equation~(\ref{deltah}) 
is then inverse-transformed to give $\Delta h(r)$. This in turn is used 
to determine $\Delta V$ by Eqs.~(\ref{energy}), (\ref{uhta}), (\ref{deltav}), 
and the 
$\Delta h(r)$ and $\Delta V$ thus obtained are used in Eqs.~(\ref{steep1}), 
(\ref{steep2}) to update $\phi(r)$ and $K$ iteratively, until convergence is
achieved. 

As observed in~\cite{pastore}, a valuable improvement over the steepest descent
is the conjugate gradient algorithm, which we have adopted here. In this 
method, the direction of descent at the step $l+1$ does not coincide with that 
of the gradient of ${\cal S}$ at step $l$, but is instead given by a linear 
combination of the gradients at all previous steps according to 
\begin{eqnarray}
\phi_{l+1}(r) & = & \phi_{l}(r) -\lambda\, \gamma_{l}(r) \mbox{\hspace{1.5cm}}
r<\sigma \, ,  
\label{conj1} \\
K^{*}_{l+1} & = & K^{*}_{l} - \lambda\, \Gamma_{l} \, , 
\label{conj2}
\end{eqnarray}
where $\gamma_{l}(r)$ and $\Gamma_{l}$ are defined by recurrence 
as~\cite{pastore}
\begin{eqnarray}
\gamma_{l}(r) & = & \Delta h_{l}(r) + \alpha_{l}\gamma_{l-1}(r) 
\mbox{\hspace{1.5cm}} r<\sigma \, ,
\label{gamma1} \\
\Gamma_{l} & = & \Delta V_{l} + \alpha_{l}\Gamma_{l-1} \, , 
\label{gamma2}
\end{eqnarray}
with $\alpha_{0}=0$ and $\alpha_{l}$ for $l\geq 1$ given by
\begin{equation}
\alpha_{l} = \frac{\rho{\displaystyle\int}_{\!\!r<1}\!\!\!d^{3}{\bf r} 
\, \Delta h_{l}(r)
\left [\Delta h_{l}(r)-\Delta h_{l-1}(r)\right] + \Delta V_{l}
\left(\Delta V_{l}-\Delta V_{l-1}\right)}{\rho{\displaystyle\int}_{\!\!r<1}
\!\!\!d^{3}{\bf r} \, 
\left[\Delta h_{l-1}(r)\right]^{2} + \left(\Delta V_{l-1}\right)^{2}} \, . 
\label{alpha}
\end{equation}
The numerator and denominator of Eq.~(\ref{alpha}) are scalar products of the
form $\langle v_{l}|v_{l}-v_{l-1}\rangle$, $\langle v_{l-1}|v_{l-1}\rangle$
for the present case in which the ``vector'' $v_{l}$ consists of both the
function $\sqrt{\rho}\Delta h_{l}(r)$ and the scalar $\Delta V$. 

In the implementation of the algorithm, special attention has been paid
to the choice of the parameter $\lambda$. In both the steepest descent and 
conjugate gradient methods, the optimal value of $\lambda$ at each step is 
determined by line minimization~\cite{pastore}. In the case of the conjugate
gradient, this amounts to minimizing 
${\cal S}[\phi_{l}(r)-\lambda\gamma_{l}(r), K^{*}_{l}-\lambda\Gamma_{l}]$ as 
a function of $\lambda$ for fixed $\phi_{l}(r)$, $\gamma_{l}(r)$, 
$K^{*}_{l}$, $\Gamma_{l}$. A possible way would be to use some method based 
on repeated evaluation of the functional ${\cal S}$ along the line 
$\phi_{l}(r)-\lambda\gamma_{l}(r)$, $K^{*}_{l}-\lambda\Gamma_{l}$. We found 
that this
procedure is not recommended here for a number of reasons: first, it is not 
very efficient, since it requires many evaluation of ${\cal S}$ at each step.
Second, at low temperature (see Sec.~\ref{sec:theory}) there is a density 
interval where the
argument of the logarithm in Eq.~(\ref{fring}) is not positive definite as a
function of $k$, so great care has to be taken to prevent the functional 
${\cal S}$ from being evaluated for these states~\cite{pastore}. Finally, there
is a subtle, yet important reason related to the fact that, for the hard-core
plus tail potential of Eq.~(\ref{pot}), both $c(r)$ and $h(r)$ are 
discontinuous at $r=\sigma$, and display additional discontinuities if $w(r)$ 
is itself discontinuous, as in the case of the SW potential. In order
to determine with good accuracy the Fourier transform of the correlation 
functions needed in the iterative procedure, it is then advisable to subtract
off the discontinuous contribution before performing the numerical Fourier 
transform, and add to the result its Fourier transform determined analytically,
as illustrated in Sec.~\ref{sec:theory}. 
However, when the functional $f_{\rm Ring}$ of Eq.~(\ref{fring}) is discretized
in the numerical calculation, the fundamental relations~(\ref{first}), 
(\ref{second}) will hold rigorously if the numerical Fourier transforms of 
$\phi(r)$ and $\Delta h(r)$ are performed on the {\em full} functions, rather
than just on their continuous part. As a consequence, the gain in accuracy 
achieved by splitting the functions to be transformed entails as a drawback that
the solution of the Euler-Lagrange equation 
$\delta f_{\rm Ring}/\delta\phi(r)=0$ for $r<\sigma$ will not rigorously be 
anymore a minimum of $f_{\rm Ring}$ as defined by Eq.~(\ref{fring}), and it
is not obvious to us whether Eq.~(\ref{fring}) can be modified so as to recover
this property. This is one more reason why we chose to perform the minimization
with respect to $\lambda$ by an algorithm which would not rely on the direct
evaluation of ${\cal S}$, and hence of $f_{\rm Ring}$. Instead, we set 
$\psi(\lambda)={\cal S}[\phi_{l}(r)-\lambda\gamma_{l}(r), 
K^{*}_{l}-\lambda\Gamma_{l}]$, and determined the line minimum by solving 
the equation
$d\psi/d\lambda=0$ with respect to $\lambda$. This was done by the 
Raphson-Newton method, namely starting
from a guess value $\lambda_{0}$ and updating $\lambda$ recursively by the
sequence
\begin{equation}
\lambda_{k+1} = -\, \frac{\left.{\displaystyle\frac{d\psi}{d\lambda}}
\right|_{\lambda_{k}}}
{\left.\displaystyle{\frac{d^{2}\psi}{d\lambda^{2}}}\right|_{\lambda_{k}}} \, .
\label{newton}
\end{equation}
In the present case we have 
\begin{eqnarray}
-\frac{d\psi}{d\lambda} & \!=\! & \int\!\!d^{3}{\bf r}\, 
\frac{\delta{\cal S}}{\delta\phi(r)}\, \gamma(r) + \frac{\partial{\cal S}}
{\partial K^{*}}\, \Gamma  \, , 
\label{num} \\
\frac{d^{2}\psi}{d\lambda^{2}} & \!=\! & \int\!\!d^{3}{\bf r}
\!\int\!\!d^{3}{\bf r}^{\prime}\, \frac{\delta^{2}{\cal S}}
{\delta\phi(r)\delta\phi(r^{\prime})}\, \gamma(r)\gamma(r^{\prime}) +
2\Gamma \!\!\int\!\!d^{3}{\bf r} \, \frac{\partial}{\partial K^{*}}
\frac{\delta{\cal S}}{\delta\phi(r)}\, \gamma(r) + 
\frac{\partial^{2}{\cal S}}{\partial {K^{*}}^{2}}\, \Gamma^{2} \, ,
\label{den}
\end{eqnarray}
where we have understood that $\gamma(r)$ is set identically to zero for 
$r>\sigma$,  
and for brevity we have omitted the index $l$ in $\phi(r)$, $\gamma(r)$, 
$K^{*}$, $\Gamma$. All the derivatives of ${\cal S}$
are evaluated at $\phi_{l}(r)-\lambda_{k}\gamma_{l}(r)$,   
$K^{*}_{l}-\lambda_{k}\Gamma_{l}$, and are easily calculated by
Eqs.~(\ref{first}), (\ref{second}), (\ref{kfirst}), (\ref{ksecond}), (\ref{s}).
The quantities required are those already needed in Eq.~(\ref{deltah}) for
$\Delta {\hat h}(k)$. 
Specifically, Eqs.~(\ref{num}), (\ref{den}) become
\begin{eqnarray}
\!\!\!\!\!-\frac{d\psi}{d\lambda} & \!\!=\!\! & \rho\!\!\int\!\!d^{3}{\bf r}\, 
\Delta h(r) \gamma(r) + \Gamma \Delta V \, ,
\label{num2} \\
\!\!\!\!\!\frac{d^{2}\psi}{d\lambda^{2}} & \!\!=\!\! & \rho\!\!
\int\!\!\frac{d^{3}{\bf k}}
{(2\pi)^{3}} \left[S(k)\hat{\gamma}(k)\right]^{2} \!-\! 2\rho\Gamma\!\!\!
\int\!\!\frac{d^{3}{\bf k}}{(2\pi)^{3}} \, S^{\, 2}(k)\hat{w}^{*}(k)
\hat{\gamma}(k)
\!+\!\rho\Gamma^{2}\!\!\!\int\!\!\frac{d^{3}{\bf k}}{(2\pi)^{3}} 
\left[S(k)\hat{w}^{*}(k)\right]^{2} \, .
\label{den2}
\end{eqnarray}
In practice, we have found that, for each conjugate gradient step 
(\ref{conj1}), (\ref{conj2}), (\ref{gamma1}), (\ref{gamma2}), a single 
Raphson-Newton step~(\ref{newton}), (\ref{num2}), (\ref{den2}) for $\lambda$ is sufficient. Iterating 
Eq.~(\ref{newton}) so as to get closer to the minimum of $\psi(\lambda)$ did 
not lead to any significant improvement in the overall speed of convergence of
the algorithm. 

We notice that at the critical point and on the spinodal curve, where the 
compressibility $\chi_{\rm red}$ diverges, 
the integrals that appear in Eq.~(\ref{den2}) diverge as well, because 
$S^{\, 2}(k)$ develops a non integrable singularity at $k=0$, see 
Eq.~(\ref{srpa}).
In order to account for this behavior, for large values of $\chi_{\rm red}$
the contribution to the integrals in the small-$k$ domain is determined 
analytically by replacing $S(k)$ with its Ornstein-Zernike form 
$S(k)\simeq 1/(\chi_{\rm red}^{-1}+b k^{2})$, and evaluating the remainder of 
the integrand at $k=0$.

\section{Numerical calculation of the hard-sphere correlation function 
at high density} 
\label{sec:appcorr}

We start from the exact relation
\begin{equation}
h(r) = c(r) + \int\!\!\frac{d^{3}{\bf k}}{(2\pi)^{3}}\, 
e^{i{\bf k}\cdot{\bf r}}\, [\hat{h}(k)-\hat{c}(k)]=
c(r) + \frac{\rho}{2\pi^{2}r}\int_{0}^{+\infty}\!\! dk \, \sin(kr)
\frac{k\, \hat{c}^{2}(k)}{1-\rho\hat{c}(k)} \, ,
\label{hs}
\end{equation}
where $\hat{c}(k)$ has been subtracted off from $\hat{h}(k)$
in order to make the integrand decrease more rapidly at large $k$.  
The function 
$\zeta(k)=k\, \hat{c}^{2}_{\rm HS}(k)/[1-\rho\hat{c}_{\rm HS}(k)]$,
which is known analytically for the Waisman parametrization, is then sampled
with a step $\Delta k\simeq 5\times 10^{-5}\sigma^{-1}$, much smaller than 
that used in
the numerical Fourier transform, so that the ``pseudo-Bragg'' peaks can be 
included in the sampling. Each peak such that the structure factor exceeds the
threshold value $S_{\rm HS}(k)=3$ is modeled by a Lorentzian 
$A/[1+\alpha^{2}(k-k_{0})^{2}]$, where the parameters $k_{0}$, $A$, $\alpha$
are determined so as to reproduce the position of the peak, its height, and its
curvature. The sum of the Lorentzians is subtracted off $\zeta(k)$, resulting 
in a smooth function that can be transformed numerically without substantial 
loss
of information, while the Fourier transform of the Lorentzians is determined 
analytically. In the present case in which $k_{0}\neq 0$ this cannot be done
exactly, but we found that for narrow peaks such that $\alpha\gg 1$ like those
considered here, a satisfactory approximation is given by
\begin{eqnarray}
\frac{1}{2\pi^{2}r} \int_{0}^{+\infty}\!\!dk \, \frac{\sin(kr)}{1+\alpha^{2}
(k-k_{0})^{2}} & \simeq & \frac{1}{2\pi\alpha r}\sin(k_{0}r)\, e^{-r/\alpha} 
\\ \nonumber
& \mbox{} - & \frac{1}{2\pi^{2}\alpha^{2}} \left\{\cos(k_{0}r){\rm Ci}(k_{0}r) 
- \sin(k_{0}r)\left[\frac{\pi}{2}-{\rm Si}(k_{0}r)\right]\right\}
\, ,
\label{lorentz}
\end{eqnarray}
where ${\rm Si}(x)$ and ${\rm Ci}(x)$ are respectively the sine and cosine 
integrals defined as:
\begin{eqnarray}
{\rm Si}(x) & = & \int_{0}^{x}\!\!dt \, \frac{\sin(t)}{t} \, , 
\label{si} \\
{\rm Ci}(x) & = & \int_{0}^{x}\!\!dt \, \frac{\cos(t)-1}{t} + \ln(x) + \gamma
\, ,
\label{ci}
\end{eqnarray} 
where $\gamma = 0.5772\ldots$ is Euler constant.

\end{document}